\documentclass[prb,aps,twocolumn,superscriptaddress,showpacs]{revtex4}
\usepackage{graphicx}
\usepackage{amsmath}
\usepackage{bm}
\newcommand{\tn}{\tilde{n}}
\newcommand{\td}{\tilde{\delta}}

\begin{document}

\author{K. Roszak}
 \affiliation{Institute of Physics, Wroc{\l}aw University of
Technology, 50-370 Wroc{\l}aw, Poland}
\author{A. Grodecka}
 \affiliation{Institute of Physics, Wroc{\l}aw University of
Technology, 50-370 Wroc{\l}aw, Poland}
\author{P. Machnikowski}
 \email{Pawel.Machnikowski@pwr.wroc.pl}
 \affiliation{Institute of Physics, Wroc{\l}aw University of
Technology, 50-370 Wroc{\l}aw, Poland}
 \affiliation{Institut f{\"u}r Festk{\"o}rpertheorie,
 Westf{\"a}lische Wilhelms-Universit{\"a}t, 48149 M{\"u}nster, Germany}
\author{T. Kuhn}
 \affiliation{Institut f{\"u}r Festk{\"o}rpertheorie,
 Westf{\"a}lische Wilhelms-Universit{\"a}t, 48149 M{\"u}nster, Germany}

\title{Phonon-induced decoherence for a quantum dot spin qubit operated
by Raman passage}

\begin{abstract}
We study single-qubit gates performed via stimulated Raman
adiabatic passage (STIRAP) on a spin qubit implemented in a
quantum dot system in the presence of phonons. We analyze the
interplay of various kinds of errors resulting from the
carrier-phonon interaction (including also the effects of 
spin-orbit coupling) as well as from quantum jumps related
to nonadiabaticity and calculate the fidelity as a function of the
pulse parameters. We give quantitative estimates for an InAs/GaAs
system and identify the parameter values for which the error is
considerably minimized, even to values below $10^{-4}$ per
operation.
\end{abstract}


\maketitle

\section{Introduction}

Quantum dots (QDs), among many other systems,\cite{bouwmeester00}
are considered to be promising candidates for an implementation of
quantum information processing schemes. Due to their atomic-like
structure \cite{jacak98a} one can easily single out a subset of
states to encode the logical qubit values. In principle, these
systems provide for stable coherent memory if the information is
encoded into the long-living electron spin,\cite{hanson03} which
motivated a spin-based proposal for quantum information storage
and processing.\cite{loss98} On the other hand, experimental
demonstrations of coherent control over the charge (orbital)
degrees of freedom
\cite{stievater01,kamada01,htoon02,zrenner02,borri02a} and the
recently performed two-qubit gate based on a confined biexciton
system \cite{li03} prove the feasibility of quantum coherent
manipulation of carrier states on picosecond time scales. It has
been therefore proposed \cite{biolatti00} to implement the qubit
states as the vacuum and single exciton states in a QD, switched
by resonant optical coupling and providing the two-cubit
conditional gating via inter-QD dipole-dipole interaction.

Both the spin-based and the charge-based proposals suffer from
serious difficulties. The spin switching time in typical
structures is very long due to weak magnetic coupling. The orbital
degrees of freedom do not provide for long operation times due to
the finite exciton lifetime, usually of order of 1
ns.\cite{borri01,bayer02} It seems therefore natural to seek for a
scheme in which the logical values are stored using spin states,
while the operations are performed via optical coupling to the
charge degrees of freedom\cite{pazy03a,calarco03}, also using QD
systems in QED cavities.\cite{imamoglu99,feng03} 

A promising
solution, proposed recently,\cite{troiani03} is to encode the
qubit states into spin states of an excess electron in a QD and
perform operations by employing the
stimulated Raman adiabatic passage (STIRAP) to a state localized
spatially in a neighboring dot.\cite{hohenester00c} (An
alternative scheme not relying on the auxiliary state has also
been proposed\cite{chen04}). The STIRAP technique uses three laser
fields that couple the two qubit states as well as the auxiliary state
to a fourth state, a charged exciton ($X^{-}$, or trion), composed of
two electrons with opposite spins and a hole. In the presence of
laser fields with slowly varying amplitudes, the system evolves
adiabatically,
following the states of the interacting system of carriers and
electromagnetic field (trapped states). The driving fields may be
chosen in such a way that the trion state is never occupied (in the ideal
case) so that the scheme is not affected by the decoherence
resulting from its finite lifetime. 
It can be shown\cite{kis02} that with properly
chosen phases of the laser pulses a pre-defined qubit superposition
gets coupled and undergoes
an adiabatic transition to the second dot and back which shifts its
phase by a desired angle with respect to the other, orthogonal
superposition that remains decoupled from the laser fields. 
This results in an arbitrary rotation of the qubit state around
an arbitrary axis on the qubit Bloch sphere.

The essential difference between the atomic systems, where such
quantum-optical schemes are successfully applied,\cite{bergmann98}
and the solid state QD systems, where their new implementation is
proposed, is the nature of the environment. In high-quality
samples at low temperatures the dominant coupling to the external
degrees of freedom is that involving lattice modes (phonons). The
coupling mechanisms include interaction with lattice polarization
(longitudinal optical, LO, phonons) and with piezoelectric fields
induced by phonon-related strain (longitudinal and transverse
acoustic, LA and TA, phonons) as well as the effective influence
of strain-induced band shift, described in terms of the
deformation potential coupling to LA phonons. Even restricted to
acoustic phonons, this kind of external bath shows various
peculiarities compared to models usually assumed in general
studies.\cite{breuer02} Its low-frequency behavior depends on the
coupling mechanism and on the wave-function geometry and is always
super-ohmic, i.e., its spectral density grows super-linearly with
frequency.\cite{krummheuer02} Due to the localization of carrier
wave-functions on a scale much larger than the lattice constant, a
high-frequency exponential cut-off in the effective phonon
spectral densities appears well below the Debye frequency. 
Moreover, apart from the non-diagonal
coupling terms describing real transitions, there is usually a
diagonal coupling which leads to pure dephasing
effects\cite{krummheuer02,vagov02a} resulting from the lattice
relaxation after a fast (compared to phonon frequencies) change of
the carrier state.\cite{vagov02a,jacak03b} Such an effect
manifests itself in optical experiments as a fast partial decay of
the signal coherence\cite{borri01,borri02a} in excellent agreement
with theoretical modeling assuming its phonon-related
origin.\cite{vagov03,vagov04}

The characteristic time scales of these intrinsically
non-Markovian pure dephasing processes are determined by the
localization (QD size) and are typically much shorter than any
real phonon-induced transition process. More importantly, they
overlap with the time scales proposed for optical qubit
control.\cite{biolatti00} It has been shown\cite{alicki04a} that
the demand to avoid these pure dephasing effects limits from
\emph{below} the gating times, thus shrinking the time scale
window defined, on the other side, by the long-time decoherence
processes (e.g. the exciton lifetime).

In this paper we study the influence of the coupling to the phonon
degrees of freedom on the fidelity of the single qubit rotation
via the STIRAP process\cite{kis02} implemented in the double-QD
structure.\cite{troiani03} Even if the possible phonon-assisted transitions
to other states may be neglected, the diagonal terms still give
rise not only to pure dephasing effects but also to transitions
between the trapped carrier-field states. The probability of these
phonon-induced transitions becomes very high if the spacing
between the trapped energy levels falls into the area of high
phonon spectral density and the overall error is roughly
proportional to the process duration. Such high error rates are
critical for quantum computation schemes where extremely high fidelity
is required (e.g., errors not higher than $\sim 10^{-4}$ per gate are
allowed for
two-qubit operations) in order to provide for scalable devices
including quantum error correction schemes. We discuss how these strong
decoherence processes may be avoided by either decreasing the
trapped level separation (low-frequency regime, exploiting the
super-ohmic behavior of spectral densities) or increasing it
beyond the cut-off (high frequency regime). We show that in both
cases one encounters a trade-off situation, due to the opposite
requirements for phonon-induced jumps (short duration) and for the
fundamental adiabaticity condition and pure dephasing (slow
operation): In the low-frequency regime, avoiding phonon-induced
transitions contradicts the condition for avoiding nonadiabatic
jumps between the trapped states, which may be overcome only by
considerably extending the process duration. In the high-frequency
case, there is a competition between the pure dephasing and the
phonon-induced transitions that is overcome by increasing the
trapped state splitting, taking advantage of the particular
structure of the phonon spectral density for a double dot
structure.

The paper is organized as follows. In the next Section \ref{sec:decoh}
we present the general derivation of the
phonon-induced error for an arbitrary system evolution.
The Section \ref{sec:model} describes the model of the specific system
discussed in the paper and derives the carrier-phonon coupling
relevant for our discussion. The Section \ref{sec:stirap} provides a
description of the STIRAP qubit rotation procedure for
completeness and necessary reference. In the central Section 
\ref{sec:phon-stirap}, the results of Sec. \ref{sec:decoh} are
applied to the STIRAP procedure described in Sec.~\ref{sec:stirap}
with the phonon perturbation derived in Sec. \ref{sec:model}. This
section contains also some general
discussion. In the Section \ref{sec:model-pulse}
we present the results for specific pulse shapes in order to get
some quantitative estimates for an InAs/GaAs QD system. 
Finally, the Section \ref{sec:concl}
summarizes and concludes the paper. In addition, some technical
details and further analysis, including the effect of the spin-orbit
coupling, are presented in the Appendices.

\section{Phonon-induced decoherence: general theory}
\label{sec:decoh}

Subject of this paper is the optically induced dynamics in a quantum dot 
structure coupled to a phonon bath.
In this Section we derive the equations for the reduced density
matrix of the carrier subsystem in the leading order in the phonon
coupling, assuming that the unperturbed (ideal) evolution of the
noninteracting system, described by the unitary evolution operator
\begin{equation}\label{unitary-evol}
    U_{0}(t)=U_{\mathrm{C}}(t)\otimes e^{-iH_{\mathrm{ph}}t/\hbar}
\end{equation}
is known (see also Ref. \onlinecite{alicki02a}).  Here,
$U_{\mathrm{C}}$ is the evolution operator for the carrier subsystem
coupled to the external light field in absence of carrier-phonon
interaction and $H_{\mathrm{ph}}$ is the free phonon Hamiltonian. The
relevant carrier states in the quantum dot are assumed to form a
discrete set $|n\rangle$, $n=0,1,2,\ldots$, and the phonons are
described by destruction and creation operators $b_{\bm{k}}$ and 
$b_{\bm{k}}^{\dag}$ referring to bulk phonon modes with wave vector 
$\bm{k}$.

The interaction between the carriers and the phonon modes 
is written in the general form
\begin{equation}\label{int-gen}
    V=\sum_{nn'}S_{nn'}\otimes R_{nn'},
\end{equation}
where $S_{nn'}=S^{\dag}_{n'n}=|n\rangle\langle n'|$ act in
the Hilbert space of the carrier subsystem and
\begin{equation}\label{Rnn}
    R_{nn'}=R_{n'n}^{\dag}
    =\sum_{\bm{k}}
    F_{nn'}(\bm{k})\left(b_{\bm{k}}+b_{\bm{-k}}^{\dag}\right),
\end{equation}
with $F_{nn'}(\bm{k})=F_{n'n}^{*}(-\bm{k})$, affect only the
phonon environment.

We assume that at the initial time $-t_{0}$ the system is in the
product state
\begin{equation}\label{init}
    \varrho(-t_{0})=\rho_{0}\otimes\rho_{T},\;\;\;
    \rho_{0}=|\psi_{0}\rangle\!\langle\psi_{0}|,
\end{equation}
where $|\psi_{0}\rangle$ is a certain state of the carrier
subsystem and $\rho_{T}$ is the thermal equilibrium distribution
of phonon modes. Physically, this is justified by the existence of
two distinct time scales: the long one for the carrier decoherence
(e.g. about 1 ns ground state exciton lifetime
\cite{borri01,bayer02}) and the short one for the reservoir
relaxation (1 ps pure dephasing time
\cite{borri01,krummheuer02,jacak03b}).

The starting point is the evolution equation for the density
matrix of the total system in the interaction picture with respect
to the externally driven evolution $U_{0}$, in the second order
Born approximation with respect to the carrier-phonon interaction
\cite{cohen98}
\begin{eqnarray}\label{evol0}
    \tilde{\varrho}(t) & = & \tilde{\varrho}(-t_{0})
    +\frac{1}{i\hbar}\int_{-t_{0}}^{t}d\tau[V(\tau),\varrho(-t_{0})] \\
\nonumber & &
-\frac{1}{\hbar^{2}}\int_{-t_{0}}^{t}d\tau\int_{-t_{0}}^{\tau}d\tau'
      [V(\tau),[V(\tau'),\varrho(-t_{0})]],
\end{eqnarray}
where
\begin{displaymath}
    \tilde{\varrho}(t)=U_{0}^{\dag}(t)\varrho(t)U_{0}(t),\;\;\;
    V(t)=U_{0}^{\dag}(t)VU_{0}(t).
\end{displaymath}

The reduced density matrix of the carrier subsystem is
\begin{displaymath}
    \rho(t)=U_{\mathrm{C}}(t)\tilde{\rho}(t)
            U_{\mathrm{C}}^{\dag}(t),\;\;\;
	\tilde{\rho}(t)=\left[
        \mathrm{Tr}_{\mathrm{R}}\tilde{\varrho}(t) \right],
\end{displaymath}
where the trace is taken over the reservoir degrees of freedom.  Note
that in this paper the symbol $\varrho$ always refers to a density
matrix in the full carrier-phonon Hilbert space while $\rho$ refers to
reduced density matrices either in the phonon or the carrier subspace.
The first (zeroth order) term in (\ref{evol0}) obviously yields
\begin{equation}\label{ro0}
    \rho^{(0)}(t)
        =U_{\mathrm{C}}(t)|\psi_{0}\rangle\!\langle\psi_{0}|
            U_{\mathrm{C}}^{\dag}(t)
        =|\psi_{0}(t)\rangle\!\langle\psi_{0}(t)|.
\end{equation}
The second term vanishes, since it contains the thermal average of
an odd number of phonon operators. The third (second order) term
describes the leading phonon correction to the dynamics of the
carrier subsystem,
\begin{eqnarray}\label{ro2}
    \lefteqn{\tilde{\rho}^{(2)}(t)=} \\
\nonumber & &
    -\frac{1}{\hbar^{2}}\int_{-t_{0}}^{t}d\tau\int_{-t_{0}}^{\tau}d\tau'
      \mathrm{Tr}_{\mathrm{R}}[V(\tau),[V(\tau'),\varrho(-t_{0})]].
\end{eqnarray}

The first of the four terms resulting from expanding the
commutators in (\ref{ro2}) is $-Q_{t}\rho_{0}$, where
\begin{eqnarray*}
    Q_{t} & = & \frac{1}{\hbar^{2}}\sum_{nn'}\sum_{mm'}
    \int_{-t_{0}}^{t}d\tau\int_{-t_{0}}^{\tau}d\tau' \\
    & & \times S_{nn'}(\tau)S_{mm'}(\tau')
     \langle R_{nn'}(\tau-\tau')R_{mm'}\rangle. \nonumber
\end{eqnarray*}
The operators $S$ and $R$ are transformed into the interaction
picture in the usual way
\begin{displaymath}
    S_{nn'}(t)=U_{0}^{\dag}(t)S_{nn'}U_{0}(t),\;\;\;
    R_{nn'}(t)=U_{0}^{\dag}(t)R_{nn'}U_{0}(t)
\end{displaymath}
and $\langle\hat{\cal O}\rangle
=\mathrm{Tr}_{\mathrm{R}}[\hat{\cal O}\rho_{T}]$ denotes the
thermal average (obviously $[U_{0}(t),\rho_{T}]=0$). Using the
symmetry of the operators $S_{nn'}$ and $R_{nn'}$ the second term
may be written as $-\rho_{0}Q^{\dag}_{t}$. In a similar manner,
the two other terms may be combined to
$\hat{\Phi}_{t}\left[\rho_{0}\right]$, where
\begin{eqnarray*}
    \hat{\Phi}_{t}\left[\rho\right] & = &
    \frac{1}{\hbar^{2}}\sum_{nn'}\sum_{mm'}
    \int_{-t_{0}}^{t}d\tau\int_{-t_{0}}^{t}d\tau' \\
  & &  \times S_{nn'}(\tau')\rho S_{mm'}(\tau)
     \langle R_{mm'}(\tau-\tau')R_{nn'}\rangle.
\end{eqnarray*}

In terms of the new hermitian operators
\begin{equation}\label{Ah}
    A_{t}=Q_{t}+Q_{t}^{\dag},\;\;\;
    h_{t}=\frac{1}{2i}(Q_{t}-Q^{\dag}_{t}),
\end{equation}
the perturbation to the density matrix at the final time $t$
(\ref{ro2}) may be written as
\begin{equation}\label{master}
    \tilde{\rho}^{(2)}(t)=
    -i\left[ h_{t},\rho_{0} \right]
    -\frac{1}{2}\left\{A_{t},\rho_{0}\right\}
    +\hat{\Phi}_{t}[\rho_{0}].
\end{equation}
The first term is a Hamiltonian correction which does not lead to
irreversible effects and in principle may be compensated for by an
appropriate modification of the control Hamiltonian
$H_{\mathrm{C}}$. The other two terms describe processes of
entangling the system with the reservoir, leading to the loss of
coherence of the carrier state.

Introducing the spectral density of the reservoir,
\begin{equation}\label{spdens}
    R_{nn',mm'}(\omega)= \frac{1}{2\pi\hbar^{2}}\int dt
    \langle R_{nn'}(t)R_{mm'}\rangle e^{i\omega t},
\end{equation}
one may write
\begin{equation}\label{Fi}
    \hat{\Phi}_{t}\left[\rho\right]=
    \sum_{nn'}\sum_{mm'}
    \int d\omega R_{nn',mm'}(\omega)
    Y_{mm'}(\omega)\rho Y_{n'n}^{\dag}(\omega)
\end{equation}
where the frequency-dependent operators have been introduced,
\begin{equation}\label{Y}
    Y_{nn'}(\omega)=\int_{-t_{0}}^{t}S_{nn'}(\tau)e^{i\omega \tau}d\tau.
\end{equation}
Using (\ref{spdens}) one has also
\begin{eqnarray*}
    Q_{t} & = &
    \sum_{nn'}\sum_{mm'}\int d\omega\int_{-t_{0}}^{t}d\tau\int_{-t_{0}}^{t}d\tau'
    \Theta(\tau-\tau')\\
    & & \times S_{nn'}(\tau)S_{mm'}(\tau')
    R_{nn',mm'}(\omega)e^{-i\omega(\tau-\tau')}.
\end{eqnarray*}
Next, representing the Heaviside function as
\begin{displaymath}
    \Theta(t)=-e^{i\omega t}\int\frac{d\omega'}{2\pi i}
    \frac{e^{-i\omega' t}}{\omega'-\omega+i0^{+}},
\end{displaymath}
we write
\begin{widetext}
\begin{eqnarray*}
    Q_{t} & = & -
    \sum_{nn'}\sum_{mm'}\int d\omega R_{nn',mm'}(\omega)
    \int\frac{d\omega'}{2\pi i}
    \frac{Y_{n'n}^{\dag}(\omega')Y_{mm'}(\omega')}{\omega'-\omega+i0^{+}} \\
    & = & -
    \sum_{nn'}\sum_{mm'}\int d\omega R_{nn',mm'}(\omega)
    \int\frac{d\omega'}{2\pi i} Y_{n'n}^{\dag}(\omega')Y_{mm'}(\omega')
    \left[-i\pi\delta(\omega'-\omega)+\mathcal{P}\frac{1}{\omega'-\omega}
    \right],
\end{eqnarray*}
\end{widetext}
where $\mathcal{P}$ denotes the principal value.

Hence, the two Hermitian operators defined in (\ref{Ah}) take the
form
\begin{equation}\label{A}
    A_{t}=
    \sum_{nn'}\sum_{mm'}\int d\omega
    R_{nn',mm'}(\omega)Y_{n'n}^{\dag}(\omega)Y_{mm'}(\omega)
\end{equation}
and
\begin{eqnarray}\label{h}
    \lefteqn{h_{t}=} \\
\nonumber
  & & \sum_{nn'}\sum_{mm'}
    \int d\omega R_{nn',mm'}(\omega)
    \mathcal{P} \int\frac{d\omega'}{2\pi}
    \frac{Y_{n'n}^{\dag}(\omega')Y_{mm'}(\omega')}{\omega'-\omega}.
\end{eqnarray}
In the following, we will be interested in the system state at the final
time $t=+t_{0}$, after all the pulses have been switched off.

In the quantum information processing context it is customary to
quantify the quality of the operation in terms of the fidelity,
which is a measure of the overlap between the desired
(unperturbed) state and the actual final state,
$F=\mathrm{Tr}[U_{\mathrm{C}}(t)\rho_{0}U_{\mathrm{C}}^{\dag}(t)
\rho(t)]$. The error is then defined as the fidelity loss,
$\delta=1-F$. From Eqs. (\ref{ro0}) and (\ref{master}) one has
\begin{eqnarray}\label{del0}
    \delta & = & -\langle\psi_{0}|\tilde{\rho}^{(2)}|\psi_{0}\rangle \\
    \nonumber
    & = & \sum_{nn'mm'}\int d\omega R_{nn',mm'}(\omega) \\
    & & \nonumber
    \times \langle \psi_{0}|Y_{n'n}^{\dag}(\omega)\mathsf{P}^{\bot}
    Y_{mm'}(\omega)|\psi_{0} \rangle,\nonumber
\end{eqnarray}
where $\mathsf{P}^{\bot}$ is the projector on the orthogonal
complement of $|\psi_{0}\rangle$ in the carrier space. In this
order the unitary correction generated by $h_{t}$ does not
contribute to the error.

The calculation presented above requires two input components: the specific
form of the interaction potential [Eq.~(\ref{int-gen})] for a given
problem and the unperturbed time evolution
[Eq.~(\ref{unitary-evol})]. These two necessary
elements are derived for our qubit system in the two following sections.

\section{The qubit system and its interaction with phonons}
\label{sec:model}

In the following part of the paper, the general theory will be applied
to a specific system of two quantum dots containing one excess
electron and coupled to the trion state in order to perform an
arbitrary rotation in the qubit space by means of the STIRAP. Here we
formulate the model of this system and derive the Hamiltonian
describing its interaction with the phonon environment.

The Hamiltonian describing this system and its coupling to lattice
modes may be written as
\begin{equation}\label{ham0}
    H=H_{\mathrm{C}}+H_{\mathrm{ph}}+V,
\end{equation}
The first term is the STIRAP Hamiltonian including both the qubit
states and the control fields. The implementation\cite{troiani03}
defines the qubit by two $\sigma_{y}$ spin eigenstates of a single
excess electron in one of the QDs (``large'') from a vertically
stacked pair.  In order to perform a general single-qubit rotation
between the two qubit states $|0\rangle$ and $|1\rangle$ an auxiliary
state $|2\rangle$ is used\cite{kis02}, in which the electron is
shifted to the second (``small'') dot and has the same spin
orientation as in $|0\rangle$. All these three states are coupled to a
fourth state $|3\rangle$, a charged exciton (trion) state, by laser
beams $\Omega_{0},\Omega_{1},\Omega_{2}$, respectively. 
The Hamiltonian for such a
system in rotating wave approximation (RWA) is
\begin{eqnarray}\label{hamC0}
    H_{\mathrm{C}} & = & \sum_{n}\epsilon_{n}|n\rangle\!\langle n|
    \\
& & +\sum_{n=0}^{2}
    \hbar\Omega_{n}(t) \left( e^{i(\omega_{n}t-\delta_{n})}
    |n\rangle\!\langle 3| +\mathrm{H.c.}\right),
\nonumber
\end{eqnarray}
where $\epsilon_{n}$ are the energies of the corresponding states, the
slowly varying pulse envelopes $\Omega_{n}(t)$ are real
and positive, $\omega_{n}$ are the corresponding laser frequencies
and and $\delta_{n}$ are the phases of the pulses.
This Hamiltonian induces the
unitary evolution described in the previous section
by the operator $U_{\mathrm{C}}$ [Eq.~(\ref{unitary-evol})].

In order to achieve the Raman coupling, the frequencies $\omega_{n}$ 
of the laser beams must be chosen such that the detunings from the
corresponding dipole transition energies $\epsilon_{3}-\epsilon_{n}$ 
are the same for all the three couplings. Therefore, we put
$\omega_{n}=\epsilon_{3}/\hbar-\epsilon_{n}/\hbar-\Delta$, $n=0,1,2$, where
$\Delta$, the common detuning, is one of the parameters to be tuned
for optimal performance. In the rotating frame, defined by $|\tn
\rangle=e^{i(\omega_{n}t-\delta_{0})}|n\rangle$,
$n=0,1,2$, the RWA Hamiltonian (\ref{hamC0}) may be written
\begin{eqnarray}\label{hamC-rot}
    \lefteqn{H_{\mathrm{C}}=} \\
\nonumber
&&  \hbar\Delta|3\rangle\!\langle 3|
    +\frac{1}{2}\sum_{n=0}^{2}
    \hbar\Omega_{n}(t)(e^{-i\td_{n}}|\tn\rangle\!\langle 3|
    +e^{i\td_{n}}|3\rangle\!\langle\tn |),
\end{eqnarray}
where $\tilde{\delta}_{n}=\delta_{n}-\delta_{0}$ (only the relative
phase of the pulses matters).

The second term describes the free phonon evolution,
\begin{displaymath}
    H_{\mathrm{ph}}=\sum\hbar\omega_{\bm{k}}
        \beta^{\dag}_{\bm{k}}\beta_{\bm{k}},
\end{displaymath}
where  $\beta^{\dag}_{\bm{k}},\beta_{\bm{k}}$ are phonon creation
and annihilation operators (with respect to the crystal ground
state). Throughout the paper, the phonon branch index will be
implicit in $\bm{k}$, unless it is explicitly written. Together with 
$H_{\mathrm{C}}$ [Eq.~(\ref{hamC0})], the above phonon Hamiltonian
describes the known, unperturbed evolution of the system, given by
Eq.~(\ref{unitary-evol}).

The final term is the carrier-phonon interaction. Since the
adiabaticity inherent in the STIRAP procedure excludes the
possibility of inducing high-frequency dynamics and also all the
trapped state splittings should be at most of several meV (to
avoid crossing with excited carrier states), the discussion will
be restricted to acoustic phonons. The Hamiltonian describing the
electron-phonon interaction in the coordinate representation is
\begin{equation}\label{int-coor}
    V=\sum_{\bm{k}}v_{\bm{k}}e^{i\bm{k}\cdot\bm{r}}
	\left(\beta_{\bm{k}}+\beta_{-\bm{k}}^{\dag}\right),
\end{equation}
where $\bm{r}$ denotes the electron coordinate (a similar contribution
appears for holes). The coupling constants for the longitudinal and transverse
phonon branches are, respectively \cite{mahan00,mahan72},
\begin{subequations}
\begin{eqnarray}\label{coupling-l}
  \lefteqn{v^{(\mathrm{l})}_{\bm{k}} =} \\
\nonumber
    && \sqrt{\frac{\hbar}{2\rho_{\mathrm{c}} V_{\mathrm{n}} 
	\omega_{\mathrm{l}}(\bm{k})}}
    \left[\sigma k
    - i \frac{de}{\varepsilon_{0}\varepsilon_{\mathrm{s}}}
        M_{\mathrm{l}}(\hat{\bm{k}}) \right],
\end{eqnarray}
and
\begin{eqnarray}
\label{coupling-t}
  \lefteqn{v^{(\mathrm{t}_{1},\mathrm{t}_{2})}_{\bm{k}} =} \\
\nonumber
    && - i \sqrt{\frac{\hbar}{2\rho_{\mathrm{c}} V_{\mathrm{n}} 
	\omega_{\mathrm{t}}(\bm{k})}}
        \frac{de}{\varepsilon_{0}\varepsilon_{\mathrm{s}}}
        M_{\mathrm{t}_{1},\mathrm{t}_{2}}(\hat{\bm{k}})
        {\cal F}_{nn'}(\bm{k}),
\end{eqnarray}
\end{subequations}
where l,$\mathrm{t}_{1}$,$\mathrm{t}_{2}$ refer to the
longitudinal and two transverse acoustic phonon branches. Here $e$
denotes the electron charge, $\rho_{\mathrm{c}}$ is the crystal density,
$V_{\mathrm{n}}$ is the normalization volume of the phonon system, 
$\omega_{\mathrm{l,t}}$ are the phonon
frequencies, $d$ is the piezoelectric constant, $\varepsilon_{0}$
is the vacuum dielectric constant, $\varepsilon_{\mathrm{s}}$ is the static
relative dielectric constant and $\sigma$ is the deformation
potential constant for electrons. The functions $M_{s}$ depend on
the orientation of the phonon wave vector \cite{mahan72}. For the
zinc-blende structure they are given by
\begin{eqnarray}\label{M}
    \lefteqn{M_{s}(\hat{\bm{k}})=}\\
\nonumber
& & 	2\left[
        \hat{k}_{x}\hat{k}_{y}(\hat{e}_{s,\bm{k}})_{z}
        +\hat{k}_{y}\hat{k}_{z}(\hat{e}_{s,\bm{k}})_{x}
        +\hat{k}_{z}\hat{k}_{x}(\hat{e}_{s,\bm{k}})_{y}
                    \right],
\end{eqnarray}
where $\hat{\bm{k}}=\bm{k}/k$ and $\hat{\bm{e}}_{s,\bm{k}}$ is the
unit polarization vector for the wave vector $\bm{k}$ and
polarization $s$.

In the basis of the confined states relevant for the STIRAP process
the carrier-phonon interaction Hamiltonian (\ref{int-coor}) reads
\begin{equation}\label{int0}
    V=\sum_{n,n'=0}^{3}
        |n\rangle\!\langle n'| \sum_{\bm{k}}
        f_{nn'}({\bm{k}})\left(\beta_{\bm{k}}
        +\beta_{-\bm{k}}^{\dag}\right),
\end{equation}
where, for single-electron states, 
$f_{nn'}(\bm{k})=v_{\bm{k}}{\cal F}_{nn'}(\bm{k})$ with 
the form factors ${\cal F}_{nn'}(\bm{k})$ depending on the wave function 
geometry and given by
\begin{equation}\label{form-general}
    {\cal F}_{nn'}(\bm{k}) =
        \int d^{3}\bm{r}
        \Psi_{n}^{*}(\bm{r})e^{i\bm{k}\cdot\bm{r}}\Psi_{n'}(\bm{r}),
\end{equation}
where $\Psi_{n}(\bm{r})$ is the envelope wave-function of the electron.
The coupling constants $f_{nn'}(\bm{k})$ include all the coupling mechanisms
relevant for a given phonon branch and have the symmetry
$f_{nn'}(\bm{k})=f_{n'n}^{*}(-\bm{k})$.

We will assume that the two spin states used to encode $|0\rangle$
and $|1\rangle$ correspond to the same orbital wave-functions so
that the couplings $f_{00}(\bm{k})$ and $f_{11}(\bm{k})$ are
equal. The couplings $f_{01}(\bm{k})$, $f_{10}(\bm{k})$,
$f_{12}(\bm{k})$ and $f_{21}(\bm{k})$ vanish since the spin
orientation in the state $|1\rangle$ differs from that in
$|0\rangle$ and $|2\rangle$ (the effects of the spin-orbit coupling
are discussed separately below).
Moreover, it is assumed that there is no overlap of wave-functions
between the states $|0\rangle$ and $|2\rangle$, so that also
$f_{02}(\bm{k})$ and $f_{20}(\bm{k})$ vanish.

An important point is that, since the electron resides normally in
the large dot, at the initial moment the lattice is relaxed to the
corresponding minimum (``dressing'' of the electron in the
coherent deformation field). This may be accounted for by defining
the modes with respect to this shifted equilibrium, so that the
ground state of the interacting system corresponds to the new
phonon vacuum, i.e., by transforming to new phonon operators
$b_{\bm{k}}$ according to
\begin{equation}\label{phon-shift}
b_{\bm{k}}=\beta_{\bm{k}}
+\frac{f_{00}^{*}(\bm{k})}{\omega_{\bm{k}}}.
\end{equation}
Upon transformation to these new modes the interaction
reads
\begin{eqnarray*}
V &  = & \sum_{n=2,3}|n\rangle\!\langle n|
    \sum_{\bm{k}} F_{nn}(\bm{k})\left(b_{\bm{k}}
    +b_{-\bm{k}}^{\dag} \right) \nonumber\\
& & +\left[\sum_{n=0}^{2}|n\rangle\!\langle 3|
    \sum_{\bm{k}} f_{n3}(\bm{k})\left( b_{\bm{k}}
    +b_{-\bm{k}}^{\dag} \right)
    +\mathrm{H.c.}\right],
\end{eqnarray*}
where $F_{nn}(\bm{k})=f_{nn}(\bm{k})-f_{00}(\bm{k})$. Moreover,
the carrier Hamiltonian undergoes a renormalization which is,
however, inessential for our discussion. In the rotating frame the
above interaction Hamiltonian reads
\begin{eqnarray}\label{int}
    V & = & \sum_{n=2,3}|\tn\rangle\!\langle \tn|
    \sum_{\bm{k}} F_{nn}(\bm{k})\left(b_{\bm{k}}
    +b_{-\bm{k}}^{\dag} \right) \\
\nonumber
    & &  \!+\left[\sum_{n=0}^{2}
    |\tn\rangle\!\langle 3|
    \sum_{\bm{k}} F_{n3}(\bm{k})\left(b_{\bm{k}}
    +b_{-\bm{k}}^{\dag} \right)
    +\mathrm{H.c.}\right].
\end{eqnarray}
where $F_{n3}(\bm{k})
=f_{n3}(\bm{k})e^{-i(\omega_{n}t-\delta_{0})}$.
This Hamiltonian is of the
form (\ref{int-gen}) with $S_{nn'}=|n\rangle\!\langle n'|$ and 
$R_{nn'}=\sum_{\bm{k}}F_{nn'}(\bm{k})(b_{\bm{k}}+b_{-\bm{k}})$. 
The spectral densities [Eq.~(\ref{spdens})] have the explicit form
\begin{eqnarray}\label{spdens-explicit}
R_{nn',mm'}(\omega) & = &
  \frac{1}{\hbar^{2}}\sum_{\bm{k}}F_{nn'}(\bm{k})F^{*}_{m'm}(\bm{k}) \\
    & & \times\left[
    (n_{\bm{k}}+1)\delta(\omega-\omega_{\bm{k}})
    +n_{\bm{k}}\delta(\omega+\omega_{\bm{k}})
    \right], \nonumber
\end{eqnarray}
where $n_{\bm{k}}$ are phonon occupation numbers.
Note that not all possible couplings appearing in the general form
of Eq.~(\ref{int-gen})
are present in our case. It is clear
from Eq. (\ref{int}) that the phonons influence the dynamics only
when a transfer from the large dot to a spatially different
carrier state (small dot or trion state) occurs.

The interaction potential given by Eq.~(\ref{int}) will be used in the
calculation of phonon-induced decoherence according to the general theory
of Sec. \ref{sec:decoh}. First, however, one has to describe the
unperturbed evolution which is the second necessary ingredient of the
calculation. This is done in the following section.

\section{The STIRAP procedure for a single-qubit rotation}
\label{sec:stirap}

In this section we present the formal description
\cite{bergmann98,kis02} of the stimulated Raman adiabatic passage
without external perturbation. Along with the results of the previous
section this will allow us to use the general theory of
Sec. \ref{sec:decoh} for the description of phonon-induced dephasing.

The system is modeled by the Hamiltonian given by
Eq.~(\ref{hamC-rot}). The envelopes of the first two pulses,
$\Omega_{0,1}$, are chosen proportional to each
other so that they may be written as
\begin{displaymath}
    \Omega_{0}(t)=\Omega_{01}(t)\cos\chi,\;\;\;
    \Omega_{1}(t)=\Omega_{01}(t)\sin\chi,
\end{displaymath}
with a certain parameter $\chi\in (0,\frac{\pi}{2})$ defining the
fixed ratio of the pulse intensities.
In terms of the new basis states
\begin{eqnarray*}
    |B\rangle & = &
    \cos\chi|\tilde{0}\rangle+e^{-i\td_{1}}\sin\chi|\tilde{1}\rangle, \\
    |D\rangle & = &
    -\sin\chi|\tilde{0}\rangle+e^{-i\td_{1}}\cos\chi|\tilde{1}\rangle,
\end{eqnarray*}
the Hamiltonian (\ref{hamC-rot}) now reads
\begin{eqnarray}\label{ham}
    H_{\mathrm{C}} & = & \hbar\Delta|3\rangle\!\langle 3|+
    \frac{\hbar}{2}\Omega_{01}(t) (|B\rangle\!\langle 3|
    +|3\rangle\!\langle B|) \\
& & +\frac{\hbar}{2}\Omega_{2}(t)
    \left(e^{-i\td_{2}}|\tilde{2}\rangle\!\langle 3|+
    e^{i\td_{2}}|3\rangle\!\langle \tilde{2}|\right). \nonumber
\end{eqnarray}
Thus, the parameters $\chi$ and $\td_{1}$ define two orthogonal states
in the qubit space. The laser 
pulses affect only one of these
states, the coupled (bright) state $|B\rangle$, while the other
orthogonal combination, $|D\rangle$, remains unaffected. 

At a
fixed time $t$, the Hamiltonian (\ref{ham}) has the eigenstates
\begin{subequations}
\begin{eqnarray}\label{states}
  |a_{0}\rangle &=&
    \cos\theta|B\rangle
    -e^{-i\td_{2}}\sin\theta|\tilde{2}\rangle \\
  |a_{-}\rangle &=& \cos\phi
        (e^{i\td_{2}}\sin\theta|B\rangle
        +\cos\theta|\tilde{2}\rangle)\\
\nonumber
    & & -e^{i\td_{2}}\sin\phi|3\rangle,\\
  |a_{+}\rangle &=& \sin\phi
        (\sin\theta|B\rangle
        +e^{-i\td_{2}}\cos\theta|\tilde{2}\rangle) \\
\nonumber
    & &  +\cos\phi|3\rangle,
\end{eqnarray}
\end{subequations}
where
\begin{displaymath}
%\begin{equation}\label{fi-teta}
        \tan\theta=\frac{\Omega_{01}}{\Omega_{2}},\;\;\;
    \sin\phi=\frac{1}{\sqrt{2}}\left(
    1-\frac{\Delta}{\sqrt{\Delta^{2}+\Omega_{01}^{2}+\Omega_{2}^{2}}}
    \right)^{1/2}.
\end{displaymath}
%\end{equation}
The corresponding eigenvalues are $\hbar\lambda_{0,\pm}$, where
\begin{equation}\label{trappedstates}
    \lambda_{0}=0,\;\;\;
    \lambda_{\pm}= \frac{\Delta \pm
    \sqrt{\Delta^{2}+\Omega_{01}^{2}+\Omega_{2}^{2}}}{2}.
\end{equation}

\begin{figure}[tb]
\begin{center}
\unitlength 1mm
\begin{picture}(60,70)(0,10)
\put(0,0){\resizebox{60mm}{!}{\includegraphics{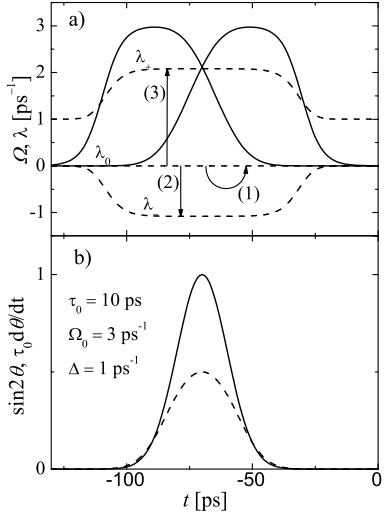}}}
\end{picture}
\end{center}
\caption{\label{fig:pulse}(a) An example of pulse shapes (solid)
and the resulting structure of the dresses levels (dashed). The arrows
show the phonon-assisted transitions, as described in
Sec. \ref{sec:phon-stirap}: (1) the pure dephasing effect, (2,3) the
transitions between the trapped states.
(b) The evolution of the functions $\dot{\theta}$ (dashed) and $\sin
2\theta$ (solid) for the pulse sequence shown in (a).}
\end{figure}

The system evolution is realized by an adiabatic change of the
pulse amplitudes (see Fig. \ref{fig:pulse}; in this application,
the detuning remains constant). Initially (at the time $-t_{0}$),
both pulses are switched off, hence $\phi=0$, then $\Omega_{2}$ is
switched on first, hence also $\theta=0$. Therefore,
$|a_{0}\rangle$ coincides with $|B\rangle$ and $|a_{-}\rangle$
with $|2\rangle$. During an adiabatic evolution of the parameters,
the states move along the corresponding spectral branches. During the
first passage, $\td_{2}=0$ and $\theta$ is changed from $0$ to
$\pi/2$. 
At the end of this stage, when the pulses are switched off ($\phi=0$),
the electron is in the state
$-|\tilde{2}\rangle$. The second passage takes $\theta$ back from
$\pi/2$ to 0. Now, however, $\td_{2}\neq 0$ so that the adiabatically
followed system state is $e^{i\td_{2}}|a_{0}\rangle$ and the final
state is $e^{i\td_{2}}|B\rangle$. Note that the desired system
evolution relies on the angle $\theta$ determined by the ratio
$\Omega_{01}/\Omega_{2}$ (so called mixing angle), 
while the absolute value of these pulse
amplitudes remains a free parameter that may be used for optimization
against decoherence effects.

The evolution operator corresponding to this procedure
may be written (in the basis $|B\rangle$,
$|\tilde{2}\rangle$,$|3\rangle$)
\begin{widetext}
%\begin{eqnarray}
\begin{equation}
\label{U0}
    U_{\mathrm{C}}(t)=
    \left( \begin{array}{ccc}
      e^{i\td_{2}}\cos\theta 
	    & e^{-i\Lambda_{-}+i\td_{2}}\cos\phi\sin\theta
            & e^{-i\Lambda_{+}}\sin\phi\sin\theta \\
      -\sin\theta
            & e^{-i\Lambda_{-}}\cos\phi\cos\theta
                 & e^{-i(\td_{2}+\Lambda_{+})}\sin\phi\cos\theta\\
      0 &   -e^{-i\Lambda_{-}+i\td_{2}}\sin\phi & e^{-i\Lambda_{+}}\cos\phi
    \end{array}\right),
\end{equation}
%\end{eqnarray}
\end{widetext}
where $\theta$,$\phi,\lambda_{\pm}$ are slowly
varying functions of time, $\td_{2}=0$ for $t<0$ and
\begin{displaymath}
    \Lambda_{\pm}(t)=\int_{0}^{t}d\tau\lambda_{\pm}(\tau).
\end{displaymath}

As shown in Ref. \onlinecite{kis02}, the phase shift of the bright
state resulting from the procedure described above is equivalent to the
rotation in the qubit space $|0\rangle,|1\rangle$ around the axis
determined by $\chi$ and by the relative phase
$\tilde{\delta}_{1}$ between $\Omega_{0}$ and $\Omega_{1}$. The
rotation angle is equal to the
$\tilde{\delta}_{2}$ phase of the second pulse
sequence. The characteristic feature of the STIRAP is that no special
form of the pulses is required. Thus, from the point of view of the
unperturbed evolution, the detuning $\Delta$ and the pulse envelopes
$\Omega_{01,2}$ are to a large extent arbitrary. This freedom may be 
used for minimizing the perturbing effects of the environment.

Ideally, the state $|2\rangle$ is only occupied during
gating, while the state $|3\rangle$ is never occupied.
This is true under the assumption that the evolution
is perfectly adiabatic. However, any change of parameters can
never be infinitely slow and the probability of a jump from
$|a_{0}\rangle$ to one of the two other states $|a_{\pm}\rangle$
remains finite. In the lowest order, the corresponding probability
amplitudes are \cite{messiah66}
\begin{displaymath}
    c_{\pm}(t)=\int_{-t_{0}}^{t}d\tau
    \langle a_{\pm}(\tau)|\dot{\psi}(\tau) \rangle
    \exp\left[ -i\int_{\tau}^{t}\lambda_{\pm}(\tau')d\tau'\right],
\end{displaymath}
where $\psi(t)$ is the state evolving adiabatically from the
initial one. Let us write the general initial state in the form
\begin{equation}\label{ini}
|\psi_{0}\rangle=\cos\frac{\vartheta}{2}|B\rangle
-e^{i\varphi}\sin\frac{\vartheta}{2}|D\rangle.
\end{equation}

The qubit rotation\cite{kis02} is performed by two well separated,
mirror-symmetric pulse sequences differing only by a phase. Thus,
using the explicit formulas (\ref{states}-c) one may write
\begin{equation}\label{jump}
    c_{\pm}(t)=
        \cos\frac{\vartheta}{2} e^{-i\Lambda_{\pm}(t)}
        \left[ \tilde{c}_{\pm} - \tilde{c}^{*}_{\pm}
        \right],
\end{equation}
where
\begin{equation}\label{c-tylda}
    \tilde{c}_{\pm}=
    \int_{-\infty}^{\infty}d\tau
    \left[\begin{array}{c}
      \sin \phi(\tau) \\
      \cos \phi(\tau) \\
    \end{array}\right]
     \dot{\theta}(\tau)
    e^{i\Lambda_{\pm}(\tau)},
\end{equation}
with $\sin \phi(\tau)$, $\cos \phi(\tau)$ corresponding to `$+$'
and `$-$', respectively, and the integral involves only one pulse
sequence. If the evolution induced by the pulse sequence is
symmetric with respect to a certain time $t_{1}$ (the time around
which the pulse sequence arrives) the amplitude
(\ref{c-tylda}) may be written in the form
$\tilde{c}_{\pm}=ie^{i\Lambda_{\pm}(t_{1})}|\tilde{c}_{\pm}|$.

In order to discuss the general properties of the nonadiabatic
jump amplitudes, let us write the evolution of $\theta$ in the
form
\begin{equation}\label{gen-teta}
    \theta(t)=\tilde{\theta}\left(\frac{t-t_{1}}{\tau_{0}}\right),
\end{equation}
where $\tilde{\theta}$ is a function of unit width, so that
$\tau_{0}$ is the time scale of the evolution of $\theta$. (The
total duration of the gate, including two pulse sequences, is
roughly an order of magnitude longer). If the functions $\phi(t)$
and $\lambda_{\pm}(t)$ change slowly around $t=t_{1}$, then
\begin{displaymath}
    |\tilde{c}_{\pm}|\approx
        \left[\begin{array}{c}
      \sin \phi(t_{1}) \\
      \cos \phi(t_{1}) \\
    \end{array}\right]
    g_{0}[\tau_{0}\lambda_{\pm}(t_{1})],
\end{displaymath}
where
\begin{displaymath}
    g_{0}(x)=\int du \tilde{\theta}'(u)e^{iux}
\end{displaymath}
is a function of unit width, with a fixed value at $x=0$ and
vanishing for $x\gg 1$ (here prime denotes the derivative with respect
to the argument $u$). Hence, the nonadiabatic jump amplitudes
(\ref{jump}) are small when
\begin{equation}\label{adiab-fund}
|\lambda_{\pm}|\gg 1/\tau_{0},
\end{equation}
which is the standard adiabaticity condition.

It is interesting to note that for symmetric pulses the final
transition probabilities (\ref{jump}) vanish for
$\Lambda_{\pm}(t_{1})=(n+1/2)\pi$. This fact is due to destructive
interference of the jump amplitudes during the first and the
second pulse sequence. Although it might be tempting to exploit
this cancellation and perform a successful passage for times and
laser beam parameters that do not satisfy the condition
(\ref{adiab-fund}), such a procedure requires a detailed knowledge
of the excitonic dipole moments and a precise control over the
laser beam properties. Moreover, the cancellation takes place only
in the final state, while during the process the other states are
occupied, which leads to the non-vanishing occupation of the
$X^{-}$ state and to decoherence, contrary to the original
motivation of this qubit implementation. In order to avoid these
effects, the envelopes of the transition probabilities should be
used as the actual bound to the nonadiabatic-jump-related error.

\section{Interaction with the phonon bath during the STIRAP
process in a QD system}
\label{sec:phon-stirap}

In this section we apply the general theory from Section
\ref{sec:decoh} to the qubit rotation performed via a STIRAP
process, as described in Section \ref{sec:stirap}, implemented in
the double-QD system.

In the Hamiltonian (\ref{int}), the only non-vanishing non-diagonal 
coupling  is $F_{n3}(\bm{k})$. Let us note, however, that for
this coupling one has, according to (\ref{Y}),
\begin{displaymath}
    Y_{n3}(\omega)=\sum_{mm'}
    \int d t e^{i[(\omega-\omega_{n}) t+\td_{n}]}
    U_{\mathrm{C}nm}^{*}U_{\mathrm{C}3m'}
    |\tilde{m}\rangle\!\langle \tilde{m}'|,
\end{displaymath}
where $U_{\mathrm{C}nm}$ are the elements of the evolution
operator (\ref{U0}),  varying at most with frequencies
$\sim\lambda_{\pm}$. It is therefore clear that this function is
peaked around $\omega\approx-\omega_{n}$, i.e., at the optical
frequencies which are many orders of magnitude higher than any
phonon frequencies present in $R_{nn',mm'}(\omega)$ [Eq.
(\ref{spdens})]. Thus, inter-band non-diagonal phonon couplings do
not contribute to (\ref{del0}). This is consistent with the
rotating wave approximation and may also be understood by noting
that the second Born approximation accounts for processes that may
be represented as a series of emission and absorption processes
involving arbitrarily many photons but only one phonon. Each
photon process takes the system from the states $0,1,2$ to $3$
with the exchange of a large energy while a non-diagonal phonon
process produces the same state change but with negligible energy
exchange. Thus, energy can never be conserved in a process
involving the inter-band phonon term.

Since the adiabatic evolution $U_{\mathrm{C}}$ does not transfer
qubit states into $|3\rangle$, $U_{\mathrm{C}}^{\dagger}|3\rangle$
remains orthogonal to $|B\rangle$. Hence, $Y_{33}(\omega)$ does
not contribute to (\ref{del0}) and we may write
\begin{equation}\label{del1}
    \delta  =
       \int d\omega \frac{R(\omega)}{\omega^{2}}S(\omega),\;\;\;
       R(\omega)\equiv R_{22,22}(\omega),
\end{equation}
with
\begin{equation}\label{s}
    S(\omega)=\omega^{2}
      \sum_{n}
      |\langle\psi_{0}|Y_{22}^{\dag}(\omega)|\psi_{n}\rangle|^{2}
      =\sum_{n}|s_{n}(\omega)|^{2},
\end{equation}
where the sum runs over a complete set of states
$|\psi_{n}\rangle$ orthogonal to $|\psi_{0}\rangle$.

For the initial state (\ref{ini}), using the explicit evolution
operator (\ref{U0}), the contributions from the three states
$|\psi_{n}\rangle =\sin\frac{1}{2}\vartheta|B\rangle
+e^{i\varphi}\cos\frac{1}{2}\vartheta|D\rangle,
|2\rangle,|3\rangle$ are, respectively,
\begin{equation}\label{s1}
  s_{1}(\omega) = -\frac{\omega}{2}\sin \vartheta
 \int_{\infty}^{\infty} dt e^{-i\omega t}\sin^{2} \theta(t),
\end{equation}
\begin{eqnarray}\label{s2}
s_{2,3}(\omega) & = & -\frac{\omega}{2}\cos\frac{\vartheta}{2}
 \int_{\infty}^{\infty} dt e^{-i\omega t}\\
&&\times   \left[\begin{array}{c}
      \cos \phi(t) \\
      \sin \phi(t) \\
    \end{array} \right]
  \sin 2\theta(t)e^{-i\Lambda_{\mp}(t)-i\td_{2}}.\nonumber
\end{eqnarray}
These three contributions correspond to transitions indicated
graphically in Fig.~\ref{fig:pulse}.

Following the argument that led to Eq. (\ref{jump}), these
functions may be written in the form
\begin{displaymath}
    |s_{n}(\omega)|=2u_{n}(\vartheta)\mathrm{Re}[\tilde{s}_{i}(\omega)],
\end{displaymath}
where $u_{1}=\frac{1}{2}\sin\vartheta$,
$u_{2,3}=\cos(\vartheta/2)$, and
\begin{subequations}
\begin{eqnarray}\label{s1a}
  \tilde{s}_{1}(\omega) &=& i\int dt e^{-i\omega t}
    \sin 2\theta(t)\dot{\theta}(t) \\
\nonumber
  & = & ie^{-i\omega t_{1}}|\tilde{s}_{1}(\omega)|,
\end{eqnarray}
\begin{eqnarray}
%  \lefteqn{
  \tilde{s}_{2,3}(\omega) & = &
  %} \\ \nonumber & &
  -\frac{\omega}{2}\int dt e^{-i[\omega t+\Lambda_{\mp}(t)-\td_{2}]}
    \left[\begin{array}{c}
      \cos \phi(t) \\
      \sin \phi(t) \\
    \end{array} \right]
    \sin 2\theta(t) \nonumber \\
\label{s23a}
    & = & e^{-i[\omega t_{1}+\Lambda_{\mp}(t_{1})+\td_{2}/2]}
    |\tilde{s}_{2,3}(\omega)|
\end{eqnarray}
\end{subequations}
where the integrals are now over one pulse sequence and the final
equalities hold for symmetric pulse sequences.

Using the representation (\ref{gen-teta}) of the system evolution
and denoting the Fourier transform of
$\tilde{\theta}'\sin2\tilde{\theta}$ by $g_{1}(x)$ we find
$|\tilde{s}_{1}(\omega)|= g_{1}(\omega\tau_{0})$. Since $t_{1}\gg
\tau_{0}$ and $|s_{1}(\omega)|^{2}$ is integrated with the slowly
varying spectral density, the oscillating terms do not contribute
and one may write
\begin{displaymath}
    |s_{1}(\omega)|^{2}\approx\frac{1}{2}\sin^{2}\vartheta
        |g_{1}(\omega\tau_{0})|^{2}.
\end{displaymath}

Hence, the function, $s_{1}(\omega)$ is centered at $\omega=0$
and broadened by a factor $1/\tau_{0}$ due to the time-dependence.
It is responsible for the pure dephasing effect.\cite{alicki04a}
The resulting error, according to (\ref{del1}), will grow with
broadening of $s_{1}(\omega)$, i.e., with decreasing process
duration. Hence, similarly to the fundamental condition
(\ref{adiab-fund}), it always favors slow operation. However, it
is independent of the trapped level splittings and reflects only
the low-frequency properties of the spectral density (at a given
temperature). For $R(\omega)\approx R_{0}\omega^{n}$, $n\ge 3$
this pure dephasing error at the temperature $T$ is
\begin{equation}\label{pure-gen}
    \delta^{(\mathrm{pd})}\sim\left\{\begin{array}{ll}
      R_{0}\tau_{0}^{-(n-1)}, & k_{\mathrm{B}}T\ll\hbar/\tau_{0} \\
      R_{0}\frac{k_{\mathrm{B}}T}{\hbar}\tau_{0}^{-(n-2)},
      & k_{\mathrm{B}}T\gg\hbar/\tau_{0}. \\
    \end{array}  \right.
\end{equation}
It should be noted that the
crossover from the low to high temperature behavior is governed only
by the pulse duration (irrespective of the system parameters) and 
for durations of the order of 10 ps it takes place at
$T\sim 0.1$ K.

By a similar argument, the two other functions may be
approximately written as
\begin{displaymath}
    |s_{2,3}(\omega)|^{2}\approx \frac{(\omega\tau_{0})^{2}}{2}
    \cos^{2}\frac{\vartheta}{2}
        \left[\begin{array}{c}
      \cos^{2} \phi \\
      \sin^{2} \phi \\
    \end{array} \right]
    g_{2}^{2}\left[(\omega+\lambda_{\mp})\tau_{0}\right],
\end{displaymath}
where $g_{2}(x)$ is the Fourier transform of $\sin
2\tilde\theta(u)$. These functions have a similar $1/\tau_{0}$
broadening but are also shifted to the spectral position
$\omega=-\lambda_{\pm}$. They describe the error resulting from
phonon-assisted transitions between the trapped states
$|a_{0,\pm}\rangle$ (see Appendix \ref{app:relax} for further
support to this interpretation). In view of the condition
(\ref{adiab-fund}), this shift must be larger than the broadening
and for rough estimates the latter may be neglected (if the
spectral density varies slowly on the scale of this broadening;
the role of oscillations in the spectral density is discussed
below). Hence, one may write
$\delta^{(\mathrm{tr})}=\delta^{(\mathrm{tr})}_{+}
+\delta^{(\mathrm{tr})}_{-}$, where
\begin{eqnarray}\label{del-tr}
    \delta^{(\mathrm{tr})}_{\mp} & = & \int d\omega
    \frac{R(\omega)}{\omega^{2}}|s_{2,3}(\omega)|^{2}\\
\nonumber
 & \approx &
    R(-\lambda_{\mp})
    \int d\omega \left|\frac{s_{2,3}(\omega)}{\omega}\right|^{2}.
\end{eqnarray}

The error is therefore proportional to
\begin{equation}\label{trans-gen}
    \delta^{\mathrm{tr}}_{\pm}\sim R(\lambda_{\pm})\tau_{0}.
\end{equation}
Thus, for a fixed spectrum of the trapped states this error grows
linearly in time (in the leading order), which is a usual
characteristics of real transition processes.

\begin{figure}[tb]
\begin{center}
\unitlength 1mm
\begin{picture}(70,90)(0,10)
\put(0,0){\resizebox{75mm}{!}{\includegraphics{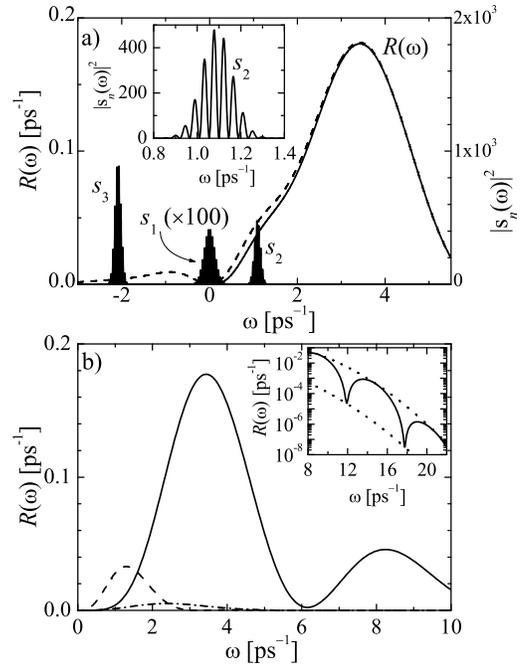}}}
\end{picture}
\end{center}
\caption{\label{fig:spdens}(a) The functions $|s_{n}(\omega)|^{2}$
describing the phonon-induced errors (for pulses as in Fig.
\ref{fig:pulse}) and the total spectral density of the phonon
reservoir $R(\omega)$ at $T=0$ (solid) and $T=5$ K (dashed) for
the model InAs/GaAs system (Tab. \ref{tab:param}). The inset shows
the exact shape of one of the spectral features. (b) The
contributions to the spectral density at T=0: DP coupling to LA
phonons (solid) and piezoelectric coupling to TA (dashed) and LA
(dash-dotted) phonons. Inset: high-frequency behavior with the two
bounds defined in Appendix \ref{app:spdens}.}
\end{figure}

In order to maximize the fidelity of the coherent operation, one
must find the trade-off between the errors caused by
phonon-assisted transitions (\ref{trans-gen}), which favors short
process durations, and the other two restrictions, related to pure
dephasing (\ref{pure-gen}) and to the general adiabaticity
condition (\ref{adiab-fund}) both increasing for fast evolution.
As can be seen from the orders of magnitude of the spectral
characteristics determining the error [Fig. \ref{fig:spdens} and
Eq. (\ref{del1})], in general, the fidelity may be strongly
decreased. However, contrary to the simple excitonic qubit case
\cite{alicki04a}, the STIRAP procedure in a QD system provides two
ways to avoid this limitations.

First, due to the super-ohmic properties of the phonon spectral
density, $R(\omega)\sim\omega^{n}$, $n\ge 2$, all error
contributions may in principle be minimized by locating the
trapped levels $\lambda_{\pm}$ in the low-frequency sector and
decreasing them while simultaneously increasing the gate duration
$\tau_{0}$.

Second, the values of $\lambda_{\pm}$ may be chosen sufficiently
far beyond the cut-off frequency. The contribution from the
phonon-induced transitions and nonadiabaticity effects may then be
arbitrarily small and the error is limited by the pure dephasing
effect, restricting the possible gate speed-up. However, one
should keep in mind that in the high frequency domain there may be
additional reservoir excitations (including two-phonon processes)
that are not accounted for in this model.

The error effects discussed here originate from the interaction
between phonons and the orbital degrees of freedom used to operate the
qubit. On the other hand, for a spin qubit one expects some
contribution to the decoherence induced by the spin-orbit (SO)
coupling. 
The electron confined in a quantum dot does not interact with other 
carriers so that, in contrast to higher-dimensional
systems \cite{dyakonov72,dyakonov86b,rossler02}, dephasing of the
electron spin
requires interaction with phonons \cite{golovach04} or nuclei
\cite{khaetskii02}. In the Appendix \ref{app:so} we analyze the former
channel. We show there that the Markovian decay of spin states in our
system is very slow and leads to negligible error over the times
relevant for the qubit operations. On the other hand, non-Markovian
SO-related effects induce the transitions between the same states
as the direct phonon coupling but are many order of magnitude weaker
due to very small SO-induced phonon coupling resulting from relatively
large energy separation of orbital states. Thus, the SO-related
effects do not affect the discussion presented here.

\section{Quantitative results for a model pulse sequence}
\label{sec:model-pulse}

In this section we calculate the errors for a STIRAP operation on
a single qubit performed with specific pulse shapes. In order to
get quantitative estimates and to identify the key error-inducing
mechanisms in various regimes of operation we use the material
parameters and QD characteristics for an InAs/GaAs system which is
frequently used as the ``typical'' system for the proposed qubit
implementations. The system parameters are collected in the Table
\ref{tab:param}.

\begin{table}
\begin{tabular}{lll}
\hline
Electron effective mass & $m^{*}$ & $0.067m_{\mathrm{e}}$\\
Static dielectric constant & $\epsilon_{\mathrm{s}}$ & 13.2 \\
Piezoelectric constant & $d$ & 0.16 C/m$^{2}$ \\
Longitudinal sound speed & $c_{\mathrm{l}}$ & 5600 m/s \\
Transverse sound speed & $c_{\mathrm{t}}$ & 2800 m/s \\
Deformation potential for electrons & $\sigma$ & $-8.0$ eV \\
Density & $\rho_{\mathrm{c}}$ & 5360 kg/m$^{3}$   \\
Land\'e factor & $g$ & -0.44 \\
Spin-orbit coupling constants: & & \\
\hspace{0.7em} Rashba & $\alpha$ & 0 \\
\hspace{0.7em} Dresselhaus & $\beta$ & 1 nm/ps \\
Level separation & $\hbar\omega_{0}$ & 46 meV \\
Electron wave-function widths: & & \\
\hspace{0.7em} in-plane & $l_{\perp}$ & 5.0 nm  \\
\hspace{0.7em} $z$-direction & $l_{z}$ & 1.5 nm  \\
Dot separation & $D$ & 6.0 nm \\
\hline
\end{tabular}
\caption{\label{tab:param}The GaAs material parameters and QD
system parameters used in the calculations 
(after Refs.~\onlinecite{adachi85,strauch90}).}
\end{table}

It is known that the STIRAP procedure is rather insensitive to the
exact pulse shape. In order to simplify the discussion, we choose
the pulse sequence
\begin{displaymath}
    \Omega_{01,2}(t) =
     \Omega_{\mathrm{env}}(t)\left[
       \frac{1\mp\sqrt{1-e^{-[(t\pm t_{1})/\tau_{0}]^{2}}}}{2}
      \right]^{1/2},
\end{displaymath}
which results in a very simple form for the time dependence of the
mixing angle,
\begin{displaymath}
    \sin 2\theta=e^{-\frac{1}{2}\left( \frac{t\pm t_{1}}{\tau_{0}}
    \right)^{2}},
    \;\;\;
    \dot{\theta}\approx\frac{1}{2\tau_{0}}e^{-\frac{1}{\pi}
        \left( \frac{t\pm t_{1}}{\tau_{0}} \right)^{2}}.
\end{displaymath}
The envelope $\Omega_{\mathrm{env}}(t)$ may be any function
approximately constant around $t_{1}$. For the numerical
calculations we take
\begin{displaymath}
    \Omega_{\mathrm{env}}(t)=
    \Omega\frac{1+\alpha}{
        1+\alpha \cosh\left( \frac{t\pm t_{1}}{\tau_{1}}\right)},
\end{displaymath}
with $\alpha=10^{-4}$, $\tau_{1}=0.4\tau_{0}$ (Fig.
\ref{fig:pulse} corresponds to this pulse choice). $\Omega^{2}$ is
proportional to the total power of the three pulses. The constant
$\Omega$, along with $\Delta$, must be tuned for
minimizing the decoherence effect.

For such a pulse sequence one finds explicitly from Eqs.
(\ref{jump},\ref{c-tylda})
\begin{eqnarray*}
    |c_{\pm}|^{2} & = & 4\pi\cos^{2}\frac{\vartheta}{2}
    \left[\begin{array}{c}
      \sin^{2} \phi \\
      \cos^{2} \phi \\
    \end{array} \right]
    \sin^{2}[\Lambda_{\pm}(t_{1})]
    e^{-2(\lambda_{\pm}\tau_{0})^{2}} \\
 & \le &  4\pi \cos^{2}\frac{\vartheta}{2}
    \left[\begin{array}{c}
      \sin^{2} \phi \\
      \cos^{2} \phi \\
    \end{array} \right]
     e^{-2(\lambda_{\pm}\tau_{0})^{2}},
\end{eqnarray*}
where the envelope of the oscillations has been taken as the safe
bound to the error, in accordance with the discussion in Sec.
\ref{sec:stirap}. For the purpose of analytical estimates the
values of $\phi=\phi(t_{1})$ and
$\lambda_{\pm}=\lambda_{\pm}(t_{1})$ are assumed constant. The
resulting error is equal to the sum of the two transition
probabilities $|c_{\pm}|^{2}$ and depends on the initial state
(\ref{ini}), since
$c_{\pm}=c_{\pm}(\vartheta,\varphi)$. The error averaged over the
initial states is
\begin{eqnarray}\label{na}
    \lefteqn{\delta^{(\mathrm{na})} =} \\
\nonumber
& & 	\frac{1}{4\pi}\int d\varphi\int d\vartheta\sin\vartheta
	\left( |c_{+}(\vartheta,\varphi)|^{2}
	      +|c_{-}(\vartheta,\varphi)|^{2} \right) \\
\nonumber
& = & 2\pi\left[
    \sin^{2}\phi e^{-2(\lambda_{-}\tau_{0})^{2}}
    +\cos^{2}\phi e^{-2(\lambda_{+}\tau_{0})^{2}} \right].
\end{eqnarray}

The spectral functions $s_{i}(\omega)$ relevant for the
phonon-induced dephasing are
\begin{equation}\label{s1-model}
    |s_{1}(\omega)|^{2}=\frac{1}{2}\sin^{2}\vartheta
    \frac{\pi^{2}}{2+\pi}\sin^{2}(\omega t_{1})
        e^{-\frac{\pi}{2+\pi}(\omega\tau_{0})^{2}}
\end{equation}
and
\begin{eqnarray}\label{s23-model}
    |s_{2,3}(\omega)|^{2} & = & 2\pi\cos^{2}\frac{\vartheta}{2}
        (\omega \tau_{0})^{2}\cos^{2}[\omega
        t_{1}+\Lambda_{\mp}(t_{1})]\\
   & & \times \left[\begin{array}{c}
      \cos^{2} \phi \\
      \sin^{2} \phi \\
    \end{array} \right]
    e^{- (\omega+\lambda_{\mp})^{2}\tau_{0}^{2}}. \nonumber
% \\
% & \approx & \pi^{3/2}\cos^{2}\frac{\vartheta}{2}
%        \left[\begin{array}{c}
%      \cos^{2} \phi \\
%      \sin^{2} \phi \\
%    \end{array} \right]
%    \tau_{0}\omega^{2} \delta(\omega+\lambda_{\mp}).
\end{eqnarray}

The total error is calculated as the sum of the nonadiabatic jump
probability (\ref{na}) and the phonon-induced contributions given
by (\ref{del1},\ref{s}) with the spectral functions
(\ref{s1-model}) and (\ref{s23-model}). The phonon spectral
density corresponding to our model double-dot InAs/GaAs system is
derived and discussed in the Appendix \ref{app:spdens} and plotted
in Fig. \ref{fig:spdens}.

\begin{figure}[tb]
\begin{center}
\unitlength 1mm
\begin{picture}(70,105)(0,5)
\put(0,0){\resizebox{70mm}{!}{\includegraphics{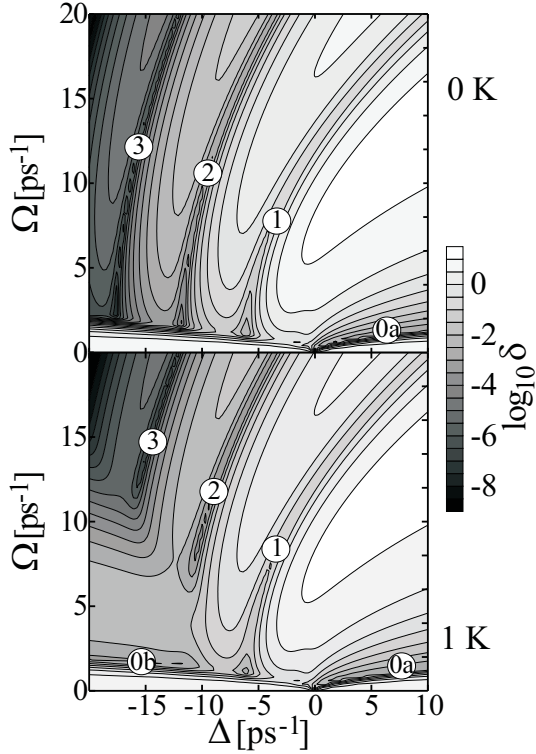}}}
\end{picture}
\end{center}
\caption{\label{fig:full-range}The dependence of $\log_{10}\delta$
on the pulse parameters $\Delta$ and $\Omega$
at $T=0$ and $T=1$ K for  $\tau_{0}=50$ ps. Numbers
refer to the parameter regimes discussed in the text.}
\end{figure}

The resulting error, averaged over $(\vartheta,\varphi)$ as in
Eq.~(\ref{na}), as a function of the pulse intensity parameter $\Omega$
and detuning $\Delta$ for a fixed process duration $\tau_{0}$ is
shown in Fig. \ref{fig:full-range}. The nontrivial interplay of
the three error contributions discussed above together with the
oscillating high-frequency tail of the phonon density of states
$R(\omega)$ (see inset in Fig. \ref{fig:spdens}b) lead to an
intricate parameter dependence of the total error. There are
clearly several parameter combinations for which the error becomes
small. With the help of the formulas (\ref{trappedstates}) one
finds that the area (0a) corresponds to $\lambda_{-}$  in the
low-frequency region, while in (0b) $\lambda_{+}$ is small and
$\lambda_{-}$ shifted beyond the phonon cut-off. The valleys
$(1),(2)\ldots$ correspond to $\lambda_{-}$ positioned at one of
the minima in the high-frequency tail of $R(\omega)$ and
$\lambda_{+}$ shifted beyond the thermal cut-off for
phonon-absorption processes, i.e. $\hbar\lambda_{+}\gtrsim
k_{\mathrm{B}}T$. For $T=0$ the absorption processes are not
allowed at all and these areas are not separated from the (0b)
region.

\begin{figure}[tb]
\begin{center}
\unitlength 1mm
\begin{picture}(60,100)(0,5)
\put(-5,0){\resizebox{70mm}{!}{\includegraphics{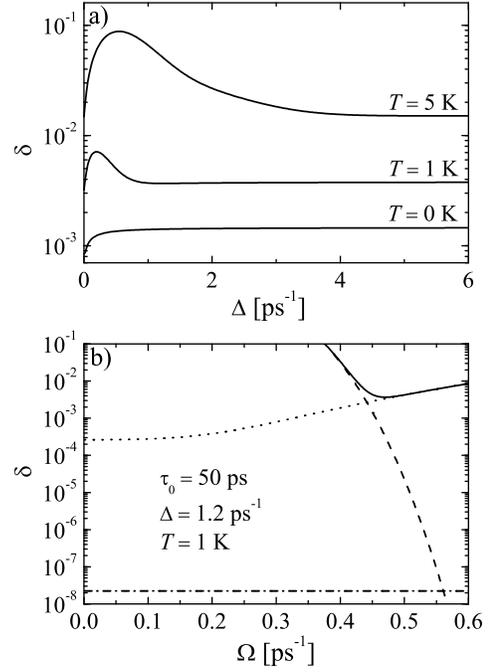}}}
\end{picture}
\end{center}
\caption{\label{fig:d-tot-low}(a) The dependence of the error for
growing detuning with $\lambda_{-}=\mathrm{const}$, along the (0a)
minimum in Fig. \ref{fig:full-range}. (b) The total error (solid)
and the individual contributions: nonadiabatic jumps (dashed),
phonon-assisted real transitions (dotted) and pure dephasing
(dash-dotted) for a section of the parameter space.}
\end{figure}

The detailed analysis of the error value along the (0a) valley at
various temperatures (Fig. \ref{fig:d-tot-low}a) shows that at
$T\neq 0$ the dependence is not monotonous. The absolute minimum
always corresponds to very low $\Omega$ and $\Delta$, for which
both trapped states $\lambda_{\pm}$ lie in the low frequency
region. At high frequencies, the error values reach a plateau
after passing (at $T>0$) through a second, very shallow minimum
(due to the subtle interplay of the error contributions weighted
by the parameter-dependent $\sin\phi$ and $\cos\phi$ factors). In
between, there is either a monotonous increase (at $T\rightarrow
0$) or a transition through a local maximum, as the $\lambda_{+}$
state crosses the frequency sector with high spectral density for
phonon absorption (cf. Fig \ref{fig:spdens}). Fig.
\ref{fig:d-tot-low}b shows the interplay between different error
contributions when the Rabi frequency $\Omega$ is changed for a
fixed detuning $\Delta$. In this range of parameters, for the
specific system under study, the pure dephasing contribution turns
out to be small compared to the errors related to real
phonon-induced transitions and to nonadiabatic jumps which create
a trade-off situation with one or two well-defined parameter sets
corresponding to the minimal errors.

\begin{figure}[tb]
\begin{center}
\unitlength 1mm
\begin{picture}(85,60)(5,8)
\put(0,0){\resizebox{95mm}{!}{\includegraphics{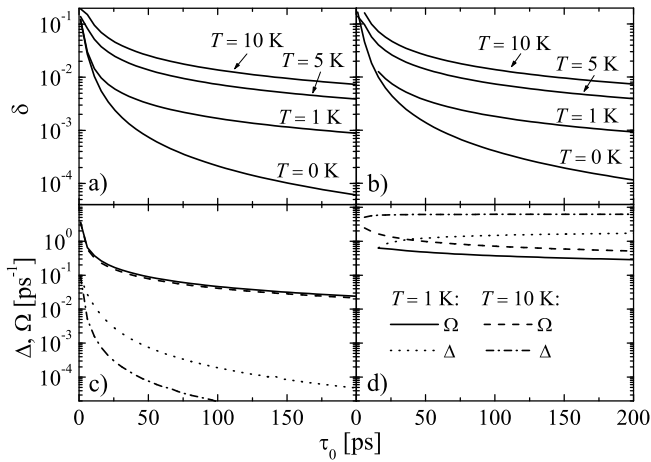}}}
\end{picture}
\end{center}
\caption{\label{fig:d-min-low}(a,b) The minimal achievable error
as a function of the process duration corresponding to the optimal
pulse parameters with both $\lambda_{\pm}$ in the low-frequency
area (a) and with $\lambda_{+}$ in the high frequency area (b)
(for $T=0$ the plateau value for $\Delta\rightarrow\infty$ is
shown). (c,d) The optimal pulse parameters (detuning and Rabi
frequency) realizing the minimal error for these two
configurations. The legend in (d) applies to both (c,d).}
\end{figure}

The above results show that for a fixed pulse duration $\tau_{0}$
the error values are bounded from below, precluding a perfect
operation for any parameter values. However, due to the
super-ohmic behavior of all the contributions to the phonon
spectral density (at low frequencies), the total error is
decreased when the process time grows and the trapped level
splittings decrease. The minimum error achievable for different
process durations at various temperatures is plotted in Fig.
\ref{fig:d-min-low}a,b and the corresponding laser beam parameters
are shown in Fig. \ref{fig:d-min-low}c,d. Both the values at the
global minimum (Fig. \ref{fig:d-min-low}a,c) and at the shallow
local minimum (Fig. \ref{fig:d-min-low}b,d) are shown. In order to
allow for any subtle interplay of parameters, for each $\tau_{0}$
the full minimization with respect to both $\Delta$ and $\Omega$
was performed. As expected, the error decreases for longer pulse
durations, but the decrease is only polynomial
($\delta\sim1/\tau_{0}$ at higher temperatures and
$\tau_{0}\gtrsim 10$ ps). Therefore, rather long pulse durations
are necessary to reduce the error considerably. Moreover, the
optimization is obtained for rather unusually small parameter
values (Fig. \ref{fig:d-min-low}b) and is very sensitive to their
precision. Still another restriction is that in this low-frequency
regime the optimum is searched against the nonadiabatic jump error
and is reached for $\tau_{0}\lambda_{\pm}\gtrsim 1$. As soon as
$\tau_{0}$ becomes comparable to the trion radiative lifetime
($\sim 1$ ns), the optimal value of $\lambda_{\pm}$ falls within
the broadening of the $|3\rangle$ state, disabling the
adiabatic passage.

The parameter dependence of the error in the (0b) area is in a way
analogous. Here, however, it is $|\lambda_{-}|$ that must be
shifted far beyond the positive frequency cut-off. Even at zero
temperature, the positive-frequency part of the spectral density
extends to relatively high frequencies (with oscillations
manifesting themselves as local minima in Fig.
\ref{fig:full-range}b). Therefore, this parameter regime is always
less favorable than the previous one.

In view of the limited possibility of fidelity optimization in the
low-frequency region for reasonable process durations, it is
interesting to study the high-frequency parameter range. In
contrast to the previous case, the values of $\Omega$ and $\Delta$
may now seem unusually high, but the results of Fig.
\ref{fig:full-range} show that by increasing the splitting between
the trapped state energies the error may in principle be reduced
to arbitrarily low values.

\begin{figure}[tb]
\begin{center}
\unitlength 1mm
\begin{picture}(85,40)(5,8)
\put(0,0){\resizebox{95mm}{!}{\includegraphics{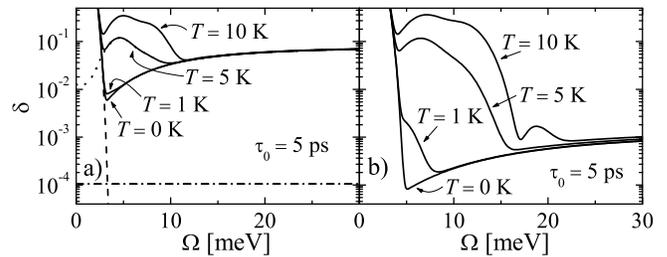}}}
\end{picture}
\end{center}
\caption{\label{fig:d-tot-high}The error as a function of $\Omega$
for $\lambda_{-}=\mathrm{const}$, along the 1 (a) and 2 (b) areas
in Fig. \ref{fig:full-range}. In (a) the individual contributions
to the error at $T=5$ K are also shown: nonadiabatic jumps
(dashed), phonon-assisted real transitions (dotted) and pure
dephasing (dash-dotted).}
\end{figure}

The Figure \ref{fig:d-tot-high} shows the error along the (1) and
(2) areas (Fig. \ref{fig:full-range}) for fixed pulse duration at
various temperatures, as well as the contributions to the error in
one case. The trapped states are now split by several meV, so that
the nonadiabatic error is negligible (except for sub-picosecond
pulses). However, the speeding-up of the dynamics is limited by
the pure-dephasing contribution. On the other hand, extending the
pulse duration is unfavorable due to the phonon-assisted
transitions. The interplay of these two contributions for a given
pulse duration, temperature, and $\lambda_{-}$ yields a series of
minima, corresponding to $\lambda_{+}$ traveling across the
oscillations of $R(\omega)$, as shown in Fig.
\ref{fig:d-tot-high}. Note that at low temperatures only one
minimum exists, belonging actually to the (0b) parameter area, but
at higher temperatures the absolute minimum shifts to the
high-frequency region.

\begin{figure}[tb]
\begin{center}
\unitlength 1mm
\begin{picture}(85,60)(5,8)
\put(0,0){\resizebox{95mm}{!}{\includegraphics{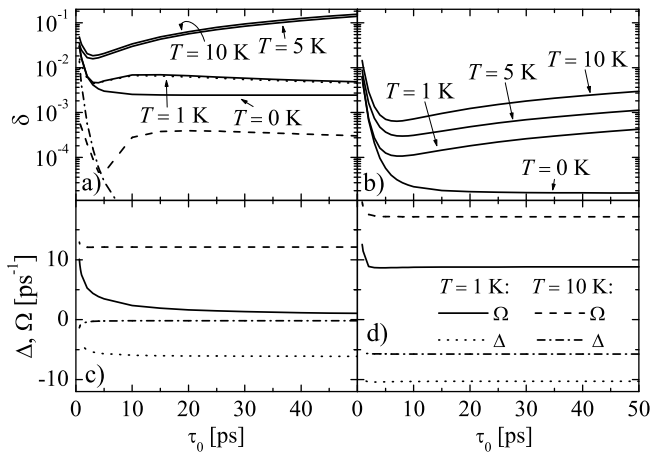}}}
\end{picture}
\end{center}
\caption{\label{fig:opty-high}(a,b) The minimal achievable error
as a function of the process duration corresponding to the optimal
pulse parameters with both $\lambda_{\pm}$ in the high-frequency
parameter areas 1 (a) and 2 (b) and the optimal pulse parameters
(detuning and Rabi frequency) realizing the minimal error for
these two configurations (c,d). In (a) the contributions to the
error at $T=1$ K are shown: nonadiabatic jumps (dashed),
phonon-assisted real transitions (dotted) and pure dephasing
(dash-dotted).}
\end{figure}

The minimum value reached depends on the pulse duration, with a
certain optimal trade-off which depends, however, on the chosen
value of $\lambda_{-}$ and decreases substantially for subsequent
minima of the spectral density. The resulting minimum value,
obtained by numerical minimization with respect to $\Omega$ and
$\Delta$ for a range of pulse durations, is shown in Fig.
\ref{fig:opty-high}a,b. The individual contributions shown in Fig.
\ref{fig:opty-high}a show that pure dephasing indeed limits the
fidelity for short pulses but in the optimal duration range the
non-monotonous $\tau_{0}$-dependence of the error is determined
exclusively by the phonon-assisted transition contribution. This
astonishing effect is in fact due to the relatively narrow minimum
of $R(\omega)$ in which $|\lambda_{-}|$ is placed. For short
pulses, $s_{3}(\omega)$ becomes broad (pure dephasing broadening
of the $\lambda_{-}$ level), increasing the overlap with
$R(\omega)$. For large $\tau_{0}$, the linear increase of $\delta$
due to long process duration becomes dominating, leading to a
minimum at a certain point.

\section{Conclusions}
\label{sec:concl}

We have studied the fidelity of the coherent operation on a QD
spin qubit rotated by a stimulated Raman adiabatic passage to a
neighboring dot and back. We have shown that, in addition to the
usual limitation of the speed of an adiabatic process, the
presence of the phonon reservoir imposes two further restrictions:
The transfer must be slow in order to minimize the pure dephasing
effect but it should not take too long in order to avoid
transitions between the trapped carrier-light states. The general
formalism was applied to an InAs/GaAs self-assembled system of
typical size. It turns out that for most values of pulse
parameters (pulse intensities and detuning) in meV range the error is
high enough to totally prevent the coherent operation. However,
there are also narrow parameter areas where the fidelity is
considerably higher.

The super-ohmic characteristics of the spectral density associated
with the phonon reservoir admits minimization of the total error
by increasing the duration of the process while simultaneously
decreasing the trapped level energies. However, the pulse
durations necessary for a considerable reduction of the error in
this low-frequency regime are of order of hundreds of picoseconds
which leads to nanosecond overall gate durations (full sequence of
two pairs of pulses). Moreover, the resulting trapped state
energies become extremely small, approaching the typical lifetime
broadening of the trion state used for the Raman coupling.

It is found that the qubit operation may be performed with much
higher fidelity if the trapped states are pushed beyond the
cut-off of the effectively coupled phonon modes. An additional
advantage comes from the oscillatory structure of the phonon
spectral density  for a double-dot system. In this way the error
at $T=0$ may be reduced to the value of $\sim 10^{-3}$ and well
below $10^{-4}$ for the trapped state energy splitting of 4 meV
and 8 meV, respectively (for the system geometry assumed here).
The latter
values lie in the spectral region where the acoustic phonon
effects dominate the decoherence, well below any spectral features
(LO phonons, higher exciton states) not included in the
discussion. It is remarkable that such low error values are
achieved with pulse durations of the order of 10 ps which,
compared to the long electron spin decoherence time, even up to tens of
miliseconds\cite{kroutvar04}, 
opens a broad time window for a large number of gating operations.

We have analyzed also the errors related to the spin-orbit
coupling. It turns out that these effects are negligible in the
present implementation.

These optimistic conclusions are somewhat limited by the strong
temperature dependence of phonon occupations, especially in the
low-frequency regime, leading to a fast increase of the error at
non-zero temperatures. Indeed, in some cases the minimal error may
grow even by an order of magnitude as soon as the temperature
reaches 1 K.

The strong dependence of the phonon-related error on the material
parameters and system geometry opens some possibility of system
engineering and optimization. For example, the high-frequency
asymptotics of the phonon spectral density is governed by the QD
height: higher dots assure a faster decay. On the other hand, in
the low-frequency sector the phonon spectral density scales with
the square of the inter-dot distance, favoring rather flat
structures. Also increasing the lateral size reduces the phonon
coupling but, at the same time, lowers the excited states
restricting the high-frequency range of operation. This shows that
finding the optimum may be nontrivial and may depend on the
frequency sector chosen for the qubit operation. It should be
noted that the high-frequency spectral density is dominated by the
deformation potential coupling which is present in any
semiconductor system but in the low-frequency domain the
piezoelectric effects dominate. This might suggest using
non-piezoelectric materials.

Let us note also that the single-qubit error calculated in this
paper gives also an estimate of the two-qubit operation if the
latter is performed using dipole coupling between the auxiliary
states in the STIRAP scheme.\cite{troiani03}

\begin{acknowledgments}
This work was supported by the Polish Ministry of Scientific
Research and Information Technology under the (solicited) Grant
No. PBZ-MIN-008/P03/2003 and by the Polish KBN under Grant No.
PB~2~P03B~08525. P.M. is grateful to the 
Humboldt Foundation for support.
\end{acknowledgments}

\appendix

\section{Phonon couplings and spectral densities}
\label{app:spdens}

In this Appendix we derive the spectral density of the phonon
reservoir $R_{22}(\omega)$ and study its properties for low and
high frequencies.

The phonon coupling constants
$F_{22}(\bm{k})=f_{22}(\bm{k})-f_{00}(\bm{k})$ have the same
structure as the original constants [Eq. (\ref{coupling-l},b)]
with the form factor replaced by ${\cal F}(\bm{k})={\cal
F}_{22}(\bm{k})-{\cal F}_{00}(\bm{k})$. Let ${\cal
F}_{\mathrm{L,S}}(\bm{k})$ denote the form factors, calculated
according to Eq. (\ref{form-general}), for the ground-state
electronic wave-function in the large (L) and small (S) dot.
Assuming that the dots are stacked along the $z$ axis at the
distance $D$, one has
\begin{displaymath}
    {\cal F}(\bm{k})
    =e^{i\frac{Dk_{z}}{2}}{\cal F}_{\mathrm{S}}(\bm{k})
    -e^{-i\frac{Dk_{z}}{2}}{\cal F}_{\mathrm{L}}(\bm{k}).
\end{displaymath}
The long-wavelength properties of the coupling constants do not
depend on the wave-function geometry. Indeed,  ${\cal
F}_{\mathrm{S,L}}(\bm{k})=1+O(k^{2})$ and ${\cal
F}(\bm{k})=iDk_{z}+O(k^{3})$.

The coupling constants for arbitrary $\bm{k}$ depend obviously on
the specific form of the wave-functions. For simplicity, we assume
Gaussian wave-functions,
\begin{displaymath}
    \Psi_{\mathrm{L,S}}(\bm{r})= N
    \exp\left[-\frac{1}{2}\left(
        \frac{\bm{r}_{\bot}}{l_{\bot\mathrm{L,S}}}\right)^{2}
    -\frac{1}{2}\left(\frac{z}{l_{z\mathrm{L,S}}}\right)\right].
\end{displaymath}
Then
\begin{equation}\label{ff-gauss}
    {\cal F}_{\mathrm{S,L}}(\bm{k})=
    e^{-\left( \frac{k_{\bot}l_{\bot\mathrm{L,S}}}{2} \right)^{2}}
    e^{-\left( \frac{k_{z}l_{z\mathrm{L,S}}}{2} \right)^{2}}.
\end{equation}
Allowing for a small difference between the dot sizes we write
$l_{\bot\mathrm{L,S}}^{2}=l_{\bot}^{2}
\pm\frac{1}{2}\Delta(l_{\bot}^{2})$,
$l_{z\mathrm{L,S}}^{2}=l_{z}^{2} \pm\frac{1}{2}\Delta(l_{z}^{2})$,
so that
\begin{eqnarray*}
    \lefteqn{{\cal F}(\bm{k})\approx} & & \\
& &  e^{-\left( \frac{k_{\bot}l_{\bot}}{2} \right)^{2}}
    e^{-\left( \frac{k_{z}l_{z}}{2} \right)^{2}} \\
& & \times\left[ 2i\sin\frac{Dk_{z}}{2}
        +\frac{k_{\bot}^{2}\Delta(l_{\bot}^{2})
            +k_{z}^{2}\Delta(l_{z}^{2})}{4}\cos\frac{Dk_{z}}{2}
          \right].
\end{eqnarray*}
Hence, the size difference brings only a small correction and will
be neglected.

Assuming isotropic phonon dispersions, the spectral density
$R(\omega)=R_{22,22}(\omega)$ [Eq.~(\ref{spdens})] may be written as
\begin{eqnarray*}
    R(\omega) & = & \frac{V}{(2\pi)^{3}}\sum_{s}\int dk k^{2} \\
 & & \times
    \left[(n_{k}+1)\delta(\omega-\omega_{k})
    +n_{k}\delta(\omega+\omega_{k}) \right] \\
& & \times\frac{1}{\hbar^{2}}\int\cos\theta d\theta \int d\varphi
    |F_{22}^{(s)}(\bm{k})|^{2}.
\end{eqnarray*}
The LA phonons are coupled both via piezoelectric and deformation
potential interaction. However, due to different inversion
symmetry the mixed terms vanish upon angle integration and the two
terms contribute independently.

The deformation potential term is
\begin{displaymath}
    R^{(\mathrm{DP})}(\omega)=
    R_{0}^{(\mathrm{DP})}\omega^{5}[n_{\mathrm{B}}(\omega)+1]
    f^{(\mathrm{DP})}(\omega),
\end{displaymath}
where
\begin{displaymath}
    R^{(\mathrm{DP})}_{0}=
    \frac{1}{3(2\pi)^{2}}\frac{D^{2}\sigma_{e}^{2}}%
        {\hbar\rho_{\mathrm{c}} c_{\mathrm{l}}^{7}}
\end{displaymath}
and the function $f^{(\mathrm{DP})}(\omega)$ is defined as
\begin{eqnarray}\label{fDP}
    f^{(\mathrm{DP})}(\omega) & = &
    \frac{3}{2}\int_{-\pi/2}^{\pi/2}d\theta\cos\theta
    \frac{4\sin^{2} \left(
    \frac{D\omega}{2c_{\mathrm{l}}}\sin\theta \right)}%
    {\left(\frac{D\omega}{2c_{\mathrm{l}}}\right)^{2}}\\
& & \times \exp\left[-\frac{1}{2}
      \left( \frac{\omega l_{\bot}}{c_{\mathrm{l}}} \right)^{2}
      \left( \cos^{2}\theta+\frac{l_{z}^{2}}{l_{\bot}^{2}}
       \sin^{2}\theta \right)\right], \nonumber
\end{eqnarray}
so that $f^{(\mathrm{DP})}(\omega)\rightarrow 1$ as
$\omega\rightarrow 0$.

For the piezoelectric contributions we choose the phonon
polarizations
\begin{eqnarray*}
    \hat{e}_{\mathrm{l},\bm{k}} &\equiv& \hat{\bm{k}}
        = (\cos\theta\cos\phi,\cos\theta\sin\phi,\sin\theta),\\
    \hat{e}_{\mathrm{t1},\bm{k}} & = & (-\sin\phi,\cos\phi,0),\\
    \hat{e}_{\mathrm{t2},\bm{k}} & = &
        (\sin\theta\cos\phi,\sin\theta\sin\phi,-\cos\theta);
\end{eqnarray*}
then the functions $M_{s}$ [Eq.~(\ref{M})] are
\begin{eqnarray*}
  M_{\mathrm{l}}  &=& \frac{3}{2}\sin2\theta\cos\theta\sin2\phi, \\
  M_{\mathrm{t1}} &=& \sin2\theta\cos2\phi, \\
  M_{\mathrm{t2}} &=& (3\sin^{2}\theta-1)\cos\theta\sin 2\phi.
\end{eqnarray*}
The corresponding terms in the spectral density are
\begin{equation}\label{sp-dens-P}
    R^{(P)}(\omega)=
    \sum_{s}R^{(\mathrm{P},s)}_{0}\omega^{3}[n_{\mathrm{B}}(\omega)+1]
    f^{(\mathrm{P},s)}(\omega),
\end{equation}
where
\begin{displaymath}
    R^{(\mathrm{P},s)}_{0}=
    \frac{1}{2\hbar\rho_{\mathrm{c}}(2\pi)^{3}c_{s}^{5}}\mu_{s}
    \left(\frac{edD}{\epsilon_{0}\epsilon_{\mathrm{s}}} \right)^{2},
\end{displaymath}
\begin{eqnarray*}
    f^{(\mathrm{P},s)}(\omega) & = &
    \frac{1}{\mu_{s}}\int_{-\pi/2}^{\pi/2}d\theta\cos\theta
    M_{s}^{2}(\theta)
    \frac{4\sin^{2} \left(
    \frac{D\omega}{2c_{\mathrm{l}}}\sin\theta \right)}%
    {\left(\frac{D\omega}{2c_{\mathrm{l}}}\right)^{2}}\\
& & \times e^{-\frac{1}{2}
      \left( \frac{\omega l_{\bot}}{c_{\mathrm{l}}} \right)^{2}
      \left[ \cos^{2}\theta+\frac{l_{z}^{2}}{l_{\bot}^{2}}
       \sin^{2}\theta \right]},
\end{eqnarray*}
$f^{(\mathrm{P},s)}(\omega)\rightarrow 1$ as $\omega\rightarrow
0$, and
\begin{displaymath}
    M_{s}^{2}(\theta)=\int d\varphi
    M_{s}^{2}(\theta,\varphi),\;\;\;
    \mu_{s}=\int d\theta\cos\theta\sin^{2}\theta
    M_{s}^{2}(\theta).
\end{displaymath}
The specific values are
$\mu_{\mathrm{l}}=\mu_{\mathrm{t}_{1}}=16\pi/35$, 
$\mu_{\mathrm{t}_{2}}=16\pi/105$.
Thus, the low-frequency behavior of the individual contributions
to the spectral density is $\sim\omega^{3}$ and $\sim\omega^{5}$
for the piezoelectric and deformation potential coupling,
respectively.

The behavior in the high frequency limit is determined by the
coupling to phonons with wave vectors in the strongest confinement
direction, i.e. along the $z$ axis. The piezoelectric coupling in
this direction is suppressed by the geometrical factors $M_{s}$
and the corresponding contributions to the spectral function
decrease rapidly. Moreover, the frequencies of TA phonons are
relatively low and the piezoelectric coupling to LA phonons is
much weaker. The frequencies of LA phonons reach much higher
values, e.g. over 20 meV for GaAs, and their dispersion remains
approximately isotropic and linear up to several
meV.\cite{strauch90} Expanding the integral into an asymptotic
series one finds an upper estimate for (\ref{fDP}),
\begin{displaymath}
    f^{(\mathrm{DP})}(\omega)\lesssim
    \frac{12c_{\mathrm{l}}^{4}}{D^{2}(l_{\bot}^{2}-l_{z}^{2})}
    \frac{1}{\omega^{4}}
    e^{-\frac{1}{2}\left( \frac{l_{z}\omega}{c_{\mathrm{l}}}
             \right)^{2}}.
\end{displaymath}
In vicinity of the points $\omega_{n}=4n\pi c_{\mathrm{l}}/D$, the
following lower bound approximately holds
\begin{equation}\label{spdens-asymp-min}
    f^{(\mathrm{DP})}(\omega)\gtrsim
    \frac{3c_{\mathrm{l}}^{6}}{(l_{\bot}^{2}-l_{z}^{2})^{3}}
    \frac{1}{\omega^{6}}
    e^{-\frac{1}{2}\left( \frac{l_{z}\omega}{c_{\mathrm{l}}}
             \right)^{2}},
\end{equation}
(see Fig. \ref{fig:spdens}b) The oscillatory behavior of the
spectral density for large frequencies follows from the fact that
the predominant contribution in this sector comes from phonons
along the strongest confinement direction, leading to a pronounced
destructive interference of interaction amplitudes in the
double-dot structure aligned along this direction.

\section{Transitions between the trapped states: Fermi Golden Rule}
\label{app:relax}

In this Appendix we show that the error
$\delta_{\mp}^{\mathrm{(tr)}}$
[Eq.~(\ref{del-tr})] may be interpreted, in terms of the Fermi Golden
Rule (FGR), as resulting from transitions between the trapped states
of the confined electron in the external driving field [transitions
(2) and (3) in Fig.~\ref{fig:pulse}].

Inserting the definition (\ref{s23a}) into the Eq. (\ref{del-tr})
and performing the frequency integration (for
$\lambda_{\pm},\phi\approx \mathrm{const}$) we get
\begin{equation}\label{del-tr-2}
    \delta^{(\mathrm{tr})}_{\mp}\approx
    \cos^{2}\frac{\vartheta}{2}
    \frac{\pi}{2} R(-\lambda_{\mp})
    \left[\begin{array}{c}
      \cos^{2} \phi \\
      \sin^{2} \phi \\
    \end{array} \right]
    \int dt \sin^{2}2\theta(t).
\end{equation}

Let us now consider the probability of phonon absorption or
emission leading to a transition from the state $|a_{0}\rangle$ to
$|a_{\pm}\rangle$. The duration of a single absorption or emission
process is of the order of the inverse phonon frequency (i.e. trapped
level spacing). Hence, in view of the adiabaticity condition
(\ref{adiab-fund}) this process is fast compared to the
characteristic time scale of the system evolution. Therefore, it
is reasonable to calculate the FGR probability
for absorption or emission at fixed values of system parameters
and include the time-dependence related to the STIRAP passage only
at the level of the rate equations. Assuming the initial state
(\ref{ini}), taking the matrix element of the phonon coupling
Hamiltonian (\ref{int}) between the trapped states
(\ref{states}-c) and applying the FGR in the standard form one
finds for the transition probability
\begin{eqnarray*}
    w_{\mp}(t) & = & \frac{2\pi}{\hbar}
    \frac{1}{4} \cos^{2}\frac{\vartheta}{2}\left[\begin{array}{c}
      \cos^{2} \phi \\
      \sin^{2} \phi \\
    \end{array} \right] \sin^{2}2\theta(t)
    \sum_{\bm{k}}|F(\bm{k})|^{2} \\
 & & \times \left[
     \delta(\hbar\lambda_{\mp}-\hbar\omega_{\bm{k}})n_{\bm{k}}
     +\delta(\hbar\lambda_{\mp}+\hbar\omega_{\bm{k}})(n_{\bm{k}}+1)
    \right] \\
 & = & \frac{\pi}{2} \cos^{2}\frac{\vartheta}{2}
    \left[\begin{array}{c}
      \cos^{2} \phi \\
      \sin^{2} \phi \\
    \end{array} \right] \sin^{2}2\theta(t) R(-\lambda_{\mp}).
\end{eqnarray*}

Solving the rate equation for the jump probability with the above
time-dependent rate $w(t)$ we find the error probability for the
whole process duration
\begin{equation}\label{del-fgr}
    \delta_{\mp}^{(\mathrm{tr})}=1-\exp\left[
    -\int_{-\infty}^{\infty}w_{\mp}(t)dt \right].
\end{equation}
For small error values this reduces to (\ref{del-tr-2}). However,
it gives also an estimate for the error beyond the applicability
of the perturbative treatment.

\section{Spin-flip effects due to spin-orbit coupling}
\label{app:so}

In the present Appendix, we discuss the additional error due to the
presence of spin-orbit coupling for the electron. We will show that
each of the spin-conserving dephasing channels discussed in the main
body of the paper is accompanied by a spin-flip channel which is,
however, several orders of magnitude weaker in a self-assembled
system. There is, moreover, an additional error related to
a spin-flip transition in the small dot but it is also
extremely small.

We start the quantitative analysis by adding the spin-orbit coupling
to the qubit Hamiltonian in Eq.~(\ref{ham0}),
\begin{displaymath}
H_{\mathrm{C}}=H_{\mathrm{d}}+H_{\mathrm{Z}}+H_{\mathrm{SO}}
	+H_{\mathrm{L}}(t).
\end{displaymath}
Here $H_{\mathrm{d}}=p^{2}/(2m^{*})+U(\bm{r})$, where $m^{*}$ is the
electron effective mass and $U(\bm{r})$ is
the confinement potential; 
$H_{\mathrm{Z}}=(1/2)g\mu_{\mathrm{B}}B\sigma_{y}$ is the
Zeeman energy ($g$ is the effective Land\'e factor, $\mu_{\mathrm{B}}$
is the Bohr magneton and $\sigma_{y}$ is the Pauli matrix; 
the magnetic field is oriented along $y$), 
$H_{\mathrm{L}}(t)$ describes the coupling to the
control laser field and 
$H_{\mathrm{SO}}=\beta(-p_{x}\sigma_{x}+p_{y}\sigma_{y})
+\alpha(p_{x}\sigma_{y}-p_{y}\sigma_{x})$ is the spin-orbit term
composed of the Rashba and Dresselhaus coupling with the constants
$\alpha$ and $\beta$, respectively. 

Following Ref. \onlinecite{golovach04}, we look for the unitary 
transformation $e^{S}$ that
eliminates the spin-orbit coupling from the stationary
Hamiltonian $H_{1}=H_{\mathrm{d}}+H_{\mathrm{Z}}+H_{\mathrm{SO}}$. To
the leading order in the SO coupling one has
\begin{displaymath}
e^{S}H_{1}e^{-S}=H_{\mathrm{d}}+H_{\mathrm{Z}}+H_{\mathrm{SO}}
+[S,H_{\mathrm{d}}+H_{\mathrm{Z}}].
\end{displaymath}
For the harmonic confinement 
$U(\bm{r})=(1/2)m^{*}\omega_{0}^{2}(x^{2}+y^{2})
+(1/2)m^{*}\omega_{z}^{2}z^{2}$, $\omega_{z}\gg\omega_{0}$, 
the SO coupling is perturbatively eliminated with the
choice
\begin{displaymath}
S=i\frac{g\mu_{\mathrm{B}}B}{(\hbar\omega)^{2}}(\beta p_{x}-\alpha
p_{y})+\ldots,
\end{displaymath}
where we omitted an irrelevant position-dependent part.

Upon the canonical transformation, the electron-phonon Hamiltonian
(\ref{int-coor}) becomes, in the leading order in the SO coupling,
$\tilde{V}=V+V_{\sigma}$, where the additional term is
\begin{eqnarray*}
V_{\sigma}& = & [S,V]\\
 & = & i\frac{g\mu_{\mathrm{B}}B}{(\hbar\omega)^{2}}
\sum_{\bm{k}}\hbar(\beta k_{x}-\alpha k_{y})v_{\bm{k}}
e^{i\bm{k}\cdot\bm{r}}(\beta_{\bm{k}}+\beta_{-\bm{k}}^{\dag})\sigma_{z},
\end{eqnarray*}
where $v_{\bm{k}}$ are defined by Eqs.~(\ref{coupling-l},b).
Within the reduced subspace spanned by the relevant states, this
operator has non-vanishing elements only between those states that have
overlapping wave functions and opposite $\sigma_{y}$ spins. Thus, one
has
\begin{eqnarray*}
V_{\sigma} & = & \sum_{\bm{k}}\left[  
F_{\sigma}^{(S)}(\bm{k}) 
(|0\rangle\!\langle 1|e^{-iE_{\mathrm{Z}}t/\hbar}+\mathrm{H.c.})\right.\\
 & & \left.+F_{\sigma}^{(L)}(\bm{k})
(|2\rangle\!\langle 2'|e^{-iE_{\mathrm{Z}}t/\hbar}+\mathrm{H.c.})
 \right](\beta_{\bm{k}}+\beta_{-\bm{k}}^{\dag}),
\end{eqnarray*}
where $E_{\mathrm{Z}}=g\mu_{\mathrm{B}}B$ is the Zeeman energy splitting,
$|2'\rangle$ is the state in the ``small'' dot with flipped spin
and
\begin{eqnarray}\label{cpl-spin}
F_{\sigma}^{(S,L)}(\bm{k}) & = & F_{\sigma}^{(S,L)*}(-\bm{k}) \\
\nonumber
& = & i\frac{g\mu_{\mathrm{B}}B}{(\hbar\omega)^{2}}\hbar
(\beta k_{x}-\alpha k_{y})v_{\bm{k}}{\cal F}_{\mathrm{S,L}}(\bm{k})
e^{\pm i\frac{k_{z}D}{2}},
\end{eqnarray}
with the form factors given by Eq.~(\ref{ff-gauss}).

Upon the phonon equilibrium shift given by Eq.~(\ref{phon-shift}), the
above interaction Hamiltonian produces a small spin-dependent 
renormalization of the qubit Hamiltonian. More importantly, the
canonical transformation implicitly performs a transition to the
eigenstates of the full Hamiltonian including the SO term. These
states may couple to the control field in a different manner than the
original states, which is reflected in the present formalism by the
correction terms resulting from the transformation 
$\tilde{H}_{L}(t)=e^{S}H_{L}(t)e^{-S}=H_{L}(t)+H_{L}^{(1)}(t)+\ldots$.
The feasibility of the STIRAP process in the presence of such
spin-dependent terms is
a separate problem, far beyond the scope of the present paper. Here we
assume that the control field can be appropriately modified so that
the new states may be evolved according to the same STIRAP transfer as
the original ones. The weakness of the spin-dependent phonon effects, 
as discussed below, suggests that also these SO corrections might be of
minor importance.

The SO contribution to the error may be written, in analogy to 
Eqs.~(\ref{del0},\ref{del1}), as
\begin{equation}\label{del-so}
\delta_{\sigma}=\int d\omega R_{\sigma}(\omega)S_{\sigma}(\omega),
\end{equation}
where
\begin{equation}\label{s-so}
S_{\sigma}(\omega)=\omega^{2}\sum_{n}
|\langle\psi_{n}|Y_{\sigma}(\omega)|\psi_{0} \rangle|^{2},
\end{equation}
with 
\begin{eqnarray*}
Y_{\sigma}(\omega)& =& \int_{-t_{0}}^{t_{0}}dt e^{i\omega t}
U_{\mathrm{C}}^{\dag}(t) \\
&&\!\!\times (|0\rangle\!\langle 1|e^{-iE_{\mathrm{Z}}t/\hbar}+
|2\rangle\!\langle 2'|e^{-iE_{\mathrm{Z}}t/\hbar}+\mathrm{H.c.})U_{\mathrm{C}}(t).
\end{eqnarray*}
The summation in Eq.~(\ref{s-so}) now involves all the states 
$|\psi_{n}\rangle$ orthogonal to $|\psi_{0}\rangle$, including $|2'\rangle$.

Similarly as in the Appendix \ref{app:spdens}, we find the
low-frequency expressions for the spectral densities
[Eq.~(\ref{spdens})]
corresponding to the SO coupling (\ref{cpl-spin}) via the two
different coupling channels
\begin{equation}\label{spdens-so-low-dp}
R_{\sigma}^{(\mathrm{DP})}(\omega)=R_{\sigma0}^{(\mathrm{DP})}
\omega^{5}[n_{\mathrm{B}}(\omega)+1],
\end{equation}
where
\begin{displaymath}
R_{\sigma0}^{(\mathrm{DP})}=
\frac{1}{12\pi^{2}}\frac{(g\mu_{\mathrm{B}}B)^{2}}{(\hbar\omega_{0})^{4}}
\frac{\hbar\sigma_{\mathrm{e}}^{2}}{\rho_{\mathrm{c}} c_{\mathrm{l}}^{7}}
(\alpha^{2}+\beta^{2}),
\end{displaymath}
and
\begin{equation}\label{spdens-so-low-p}
R_{\sigma}^{(\mathrm{P},s)}(\omega)=R_{\sigma0}^{(\mathrm{P},s)}
\omega^{3}[n_{\mathrm{B}}(\omega)+1],
\end{equation}
where 
\begin{displaymath}
R_{\sigma0}^{(\mathrm{P},s)}=
\frac{\gamma_{s}}{\pi^{2}}
\frac{(g\mu_{\mathrm{B}}B)^{2}}{(\hbar\omega_{0})^{4}}
\frac{\hbar}{\rho_{\mathrm{c}} c_{s}^{5}}
\left( \frac{de}{\epsilon_{0}\epsilon_{\mathrm{s}}} \right)^{2}
(\alpha^{2}+\beta^{2}),
\end{displaymath}
with $\gamma_{\mathrm{l}}=1/35$, 
$\gamma_{\mathrm{t}_{1}}=\gamma_{\mathrm{t}_{2}}=2/105$.
In the frequency range typical for the Zeeman energies in GaAs at
moderate magnetic fields ($\sim 0.1$ meV) the piezoelectric coupling to
transverse modes dominates.

In the high-frequency region, where the deformation potential coupling
dominates, we find the asymptotic estimate
\begin{eqnarray}\label{spdens-so-high-dp}
R_{\sigma}^{(\mathrm{DP})}(\omega)& = & 
\frac{1}{4\pi^{2}}\frac{(g\mu_{\mathrm{B}}B)^{2}}{(\hbar\omega_{0})^{4}}
\frac{\hbar\sigma_{\mathrm{e}}^{2}}{\rho_{\mathrm{c}} c_{\mathrm{l}}^{3}}
(\alpha^{2}+\beta^{2})\\
\nonumber
&&\times [n_{\mathrm{B}}(\omega)+1]
e^{-\frac{1}{2}(\frac{\omega l_{z}}{c_{\mathrm{l}}})^{2}}
\frac{\omega}{(l_{\perp}^{2}-l_{z}^{2})^{2}}.
\end{eqnarray}

The spectral function $S_{\sigma}(\omega)$, pertaining
to the driven evolution of the system, depends in a complicated way on the
performed qubit rotation and on the initial state. In order to reduce
the complexity, we restrict the discussion to a $\pi/2$ qubit rotation
around the $\sigma_{x}$ axis, i.e. $\chi=\pi/4$,
$\delta_{1}=0$. We parameterize the general initial qubit state
in the form (\ref{ini}) and calculate the spin-flip contributions to
the error averaged over the Bloch sphere of initial states.
We will restrict the
discussion to the most interesting high-frequency regime of operation,
where the following hierarchy of time scales may be assumed: $t_{0}\gg
\hbar/E_{\mathrm{Z}}\gg t_{1}\gg\tau_{0}$.

Under these simplifying assumptions and neglecting terms
proportional to $E_{\mathrm{Z}}\tau_{0}$ the first contribution to the
spectral function $S_{\sigma}(\omega)$ [Eq.~(\ref{s-so})] is
\begin{eqnarray*}
\lefteqn{|\langle 0|Y_{\sigma}(\omega)|1\rangle|^{2} \approx}\\
&& \frac{2\pi}{3}(2t_{0}-2t_{1})\left[ 
  \delta(\omega+E_{\mathrm{Z}}/\hbar)+\delta(\omega-E_{\mathrm{Z}}/\hbar)
 \right] \\
&&+\frac{1}{\omega^{2}}|s_{1}(\omega)|^{2}_{\mathrm{av}},
\end{eqnarray*}
where `av' denotes averaging over the Bloch sphere of initial states,
as in Eq.~(\ref{na}).
The first contribution leads to the Markovian (Fermi Golden Rule)
spin-flip probability
over the time $2t_{0}-2t_{1}$ during which the electron is located in the
first (``large'') dot, up to the factor 1/3 resulting from averaging of the
spin-flip rates for various superposition states. The probability of
such a process in a self-assembled quantum dot is extremely low due to
large confinement energy. Indeed, the spin-flip rate pertaining to
this contribution is, according to Eq.~(\ref{del-so}),
\begin{eqnarray}\label{FGR-spin-flip}
w & = & \frac{2\pi}{3}\left[ 
 R_{\sigma}(E_{\mathrm{Z}}/\hbar)+R_{\sigma}(-E_{\mathrm{Z}}/\hbar)
	 \right] \\
\nonumber
& \approx & \frac{1}{3}\frac{16}{105\pi}
 \left( \frac{g\mu_{\mathrm{B}}B}{\hbar\omega_{0}} \right)^{4}
 k_{\mathrm{B}}T \frac{1}{\hbar^{2}\rho_{\mathrm{c}} c_{\mathrm{t}}^{2}}
 \left( \frac{de}{\varepsilon_{0}\varepsilon_{\mathrm{s}}} \right)^{2}
 (\alpha^{2}+\beta^{2})\\
\nonumber
& = & 1.0\cdot 10^{-4}\;\mathrm{s}^{-1},  
\end{eqnarray}
where we used the low-frequency formula (\ref{spdens-so-low-p})
for the spectral density (the piezoelectric coupling to transverse
phonons dominates in this sector), substituted the Zeeman energy 
$E_{\mathrm{Z}}=g\mu_{\mathrm{B}}B$ and assumed that 
$k_{\mathrm{B}}T>E_{\mathrm{Z}}$. The final value corresponds to $T=1$
K and $B=1$ T. 

The other term describes an additional contribution to the
spin-flip transition closely
related to the pure dephasing effect [Eq.~(\ref{pure-gen})]. Since the
low-frequency behavior of the spectral densities for direct
and SO-induced processes [Eqs.~(\ref{sp-dens-P}) and
(\ref{spdens-so-low-p})] is the same, the ratio between the SO
effect and the direct pure dephasing is
\begin{displaymath}
\frac{\delta_{\sigma}^{(1)}}{\delta^{(\mathrm{pd})}}=
\frac{R_{\sigma 0}^{(\mathrm{P,t})}}{R_{0}^{(\mathrm{P,t})}}=
\frac{\hbar^{2}(\alpha^{2}+\beta^{2})}{D^{2}}
\frac{(g\mu_{\mathrm{B}}B)^{2}}{(\hbar\omega_{0})^{4}}
=1.7\cdot 10^{-12},
\end{displaymath}
where we included again only the dominating contribution from the
piezoelectric coupling to transverse phonons. It is clear that the
SO-related process is negligible compared to the pure dephasing.

Under the same assumptions as above, we have for the two other
SO-induced contributions
\begin{displaymath}
|\langle 2,3|Y(\omega)|\psi_{0}\rangle|_{\mathrm{av}}^{2}=
\frac{1}{\omega^{2}}|s_{2,3}(\omega)|^{2}.
\end{displaymath}
As discussed in Sec. \ref{sec:phon-stirap}, the functions
$s_{2,3}(\omega)$ are relatively sharply peaked around $\omega=-
\lambda_{\mp}$. Since the latter values are large, we use the high-frequency
asymptotics for the spectral density (now the deformation potential
coupling to longitudinal phonons dominates) and the ratio between the
SO-related process and the transitions described in
Sec. \ref{sec:phon-stirap} may be written as
\begin{eqnarray*}
\frac{\delta_{\sigma}^{(2,3)}}{\delta^{(2,3)}} & = & 
\frac{R_{\sigma}^{(\mathrm{DP})}(-\lambda_{\mp})}%
{R^{(\mathrm{DP})}(-\lambda_{\mp})}\\
 & = & \frac{(g\mu_{\mathrm{B}}B)^{2}}{(\hbar\omega_{0})^{4}}
\frac{\hbar^{2}}{D^{2}c_{\mathrm{l}}^{2}}\lambda_{\mp}^{2}
(l_{\perp}^{2}-l_{z}^{2})(\alpha^{2}+\beta^{2}),
\end{eqnarray*}
where we used the asymptotic formula (\ref{spdens-asymp-min}) since
the optimal pulse parameters correspond to $\lambda_{\mp}$
in a local minimum of the spectral density.
For the parameter area 1 [cf. Fig.~\ref{fig:full-range}], i.e. 
$\lambda_{-}=12$ ps$^{-1}$, one finds
\begin{displaymath}
\frac{\delta_{\sigma}^{(2)}}{\delta^{(2)}}=1.8\cdot 10^{-10}.
\end{displaymath}
Again, the SO-related process is many orders of magnitude weaker than
the transition discussed in the paper.

Apart from these contributions there is another one, to the
spin-flipped state $|2'\rangle$ in the small dot. This process is not
possible in the absence of the SO coupling. The relevant spectral
function is
\begin{eqnarray*}
\lefteqn{|\langle 2'|Y(\omega)|\psi_{0}\rangle|_{\mathrm{av}}^{2}=}\\
&&2\frac{\sin^{2}[(\omega-E_{\mathrm{Z}}/\hbar)t_{1}+g(\omega)]}%
{(\omega-E_\mathrm{Z}/\hbar)^{2}} h^{2}(\omega-E_{\mathrm{Z}}/\hbar),
\end{eqnarray*}
where we denoted
\begin{displaymath}
\int dt \cos\theta\dot{\theta}e^{i\omega t}=e^{ig(\omega)}h(\omega).
\end{displaymath}
One has for the STIRAP transfer $g(0)=0$, $h(0)=-1$. The width of
$h(\omega)$ is of the order of $1/\tau_{0}$. For large $t_{1}$ (long
dwelling time in the small dot), this yields the spin-flip transition
rate according to the Fermi Golden Rule, analogous to
Eq.~(\ref{FGR-spin-flip}) up to an averaging-related factor. For
typical dwelling times $\sim 100$ ps this would produce a negligible 
contribution
to the error of order of $10^{-14}$ for $B\sim 1$ T, $T\sim 1$ K. 
However, under the assumptions made above the FGR is not applicable; instead
one may approximate
\begin{displaymath}
|\langle 2'|Y(\omega)|\psi_{0}\rangle|_{\mathrm{av}}^{2}\approx
\frac{1}{\omega^{2}}h^{2}(\omega).
\end{displaymath}
Using the low-frequency and high-temperature approximation to 
$R_{\sigma}^{(\mathrm{P,t})}(\omega)$ one finds
\begin{displaymath}
\delta^{(2')}_{\sigma}\sim 
\left[ 
R_{\sigma 0}^{(\mathrm{P,t_{1}})}+R_{\sigma 0}^{(\mathrm{P,t_{2}})}
\right]
\frac{k_{\mathrm{B}}T}{\hbar}\frac{1}{\tau_{0}}
\sim 10^{-14},
\end{displaymath}
at $B\sim 1$ T, $T\sim 1$ K. Note that under the conditions assumed
here, the dominating contribution comes from the dynamical
effect characterized by the inverse pulse duration $1/\tau_{0}$.

In conclusion, the Markovian spin-flip rate for a self-assembled QD
with typical level separation is very long while dynamical effects
involving the SO coupling remain in a fixed relation to those induced
by direct phonon coupling and are negligible in comparison to them.

\bibliographystyle{prsty}
\bibliography{abbr,quantum}

\end{document}